\documentclass[conference]{IEEEtran}
\usepackage{fancyhdr}
\IEEEoverridecommandlockouts

\usepackage{siunitx}
\def\BibTeX{{\rm B\kern-.05em{\sc i\kern-.025em b}\kern-.08emT\kern-.1667em\lower.7ex\hbox{E}\kern-.125emX}}
    
\usepackage[belowskip=-10pt,aboveskip=2pt]{caption}
\usepackage[labelformat=simple,skip=0pt]{subcaption}

\usepackage[noadjust]{cite}
\usepackage{amsmath,amssymb,amsfonts}
\usepackage{algorithmic}
\usepackage{graphicx}
\usepackage{textcomp}
\usepackage{xcolor}
\usepackage{soul}
\usepackage{verbatim}
\usepackage{siunitx}

\usepackage{algorithm2e}
\usepackage{listings}
\definecolor{codered}{rgb}{0.75,0.0,0.0}
\definecolor{codegreen}{rgb}{0.0,0.4,0.0}
\definecolor{codeblue}{rgb}{0.0,0.0,0.6}
\usepackage{url}
\usepackage{xcolor}
\definecolor{light-gray}{gray}{0.95}

\makeatletter
\newcommand\notsotiny{\@setfontsize\notsotiny\@vipt\@viipt}
\makeatother
\usepackage[font=small]{caption}

\usepackage{booktabs,caption}
\usepackage[flushleft]{threeparttable}

\usepackage{afterpage}
\usepackage{multirow}
\usepackage{multicol}

\newcommand{\ignore}[1]{}
\newif\ifsubmit
\submitfalse
\ifsubmit
    \newcommand{\mert}[1]{}
    \newcommand{\vikram}[1]{}
    \newcommand{\tekin}[1]{}
    \newcommand{\doga}[1]{}
    \newcommand{\simon}[1]{}
    \newcommand{\bin}[1]{}
    \newcommand{\daniel}[1]{}
    \newcommand{\weng}[1]{}
    
    \newcommand{\wenmei}[1]{}
    \newcommand{\raj}[1]{}
    \newcommand{\ian}[1]{}
\else
    \definecolor{gray}{rgb}{0.66, 0.66, 0.66}
    \definecolor{dgreen}{rgb}{0.00, 0.75, 0.00}
    \definecolor{dblue}{rgb}{0.00, 0.00, 0.75}
    \newcommand{\mert}[1]{[{\color{gray}SH: #1}]}
    \newcommand{\vikram}[1]{[{\color{red}VK: #1}]}
    \newcommand{\tekin}[1]{[{\color{dblue}tekin: #1}]}
    \newcommand{\doga}[1]{{\color{violet}#1}}
    \newcommand{\simon}[1]{{\color{red}#1}}
    \newcommand{\bin}[1]{[{\color{gray}SG: #1}]}
    \newcommand{\weng}[1]{[{\color{gray}WMH: #1}]}
    \newcommand{\wenmei}[1]{[{\color{dgreen}ASSIGN: #1}]}
    \newcommand{\raj}[1]{[{\color{violet}RK: #1}]}
    \newcommand{\ian}[1]{[{\color{red}Ian: #1}]}    \newcommand{\todo}[1]{[{\color{red}TODO: #1}]}
\fi

\newcommand{\reb}[1]{{#1}}

\author{\IEEEauthorblockN{Mert Hidayeto\u{g}lu, Tekin Bicer\IEEEauthorrefmark{2}, Simon Garcia de Gonzalo\IEEEauthorrefmark{3}, Bin Ren\IEEEauthorrefmark{4}, Vincent De Andrade\IEEEauthorrefmark{2},\\ Doga Gursoy\IEEEauthorrefmark{2}, Raj Kettimuthu\IEEEauthorrefmark{2}, Ian T. Foster\IEEEauthorrefmark{2}, and Wen-mei W. Hwu\vspace{0.1cm}}

\IEEEauthorblockA{
University of Illinois at Urbana-Champaign, USA\\
\IEEEauthorrefmark{2}Argonne National Laboratory, IL, USA,\\
\IEEEauthorrefmark{3}Barcelona Supercomputing Center, Spain\\
\IEEEauthorrefmark{4}College of William \& Mary, VA, USA
\vspace{-0.5cm}}

}

\usepackage{flushend}
\usepackage{enumitem}

\begin{document}

\title{ Petascale XCT: 3D Image Reconstruction with Hierarchical Communications on Multi-GPU Nodes
}

\maketitle
\thispagestyle{fancy}
\lhead{}
\rhead{}
\chead{}
\lfoot{\footnotesize{
SC20, November 9-19, 2020, Is Everywhere We Are
\newline 978-1-7281-9998-6/20/\$31.00 \copyright 2020 IEEE}}
\rfoot{}
\cfoot{}
\renewcommand{\headrulewidth}{0pt}
\renewcommand{\footrulewidth}{0pt}

\begin{abstract}
X-ray computed tomography is a commonly used technique for noninvasive imaging at synchrotron facilities. 
Iterative tomographic reconstruction algorithms are often preferred for recovering high quality 3D volumetric images from 2D X-ray images, however, their use has been limited to small/medium datasets due to their computational requirements. 
In this paper, we propose a high-performance iterative reconstruction system for terabyte(s)-scale 3D volumes. 
Our design involves three novel optimizations: 
(1) optimization of (back)projection operators by extending \reb{the} 2D memory-centric approach to 3D; 
(2) performing hierarchical communications by exploiting ``fat-node" architecture with many GPUs; 
(3) utilization of mixed\reb{-}precision types while preserving convergence rate and quality. 
We extensively evaluate the proposed optimizations and scaling on the Summit supercomputer. 
Our largest reconstruction is a mouse brain volume with 9K$\times$11K$\times$11K voxels, where the total reconstruction time is under three minutes using 24,576 GPUs, reaching 65 PFLOPS: \reb{34\% of Summit's peak performance}.

\end{abstract}




\section{Introduction}
Synchrotron light source facilities around the world help tens of thousands of researchers every year carry out extremely challenging experiments and ground-breaking research.
X-ray computed tomography (XCT) is one of the widely used imaging modalities at synchrotron light sources for imaging materials, samples and biological specimens in 3D with high temporal and spatial resolution, ranging from several micrometers down to sub-\SI{20} nanometer resolution. 
The 
high-energy synchrotron X-ray sources, such as the Advanced Photon Source (APS) at Argonne National Laboratory (ANL), enable imaging thick specimens that can yield massive amounts of measurements, exceeding tens of GB/s rates and producing
TBs-scale data per experiment~\cite{aps-science-2018}.
For example, imaging a single adult mouse brain of a few centimeters diameter at \SI{}{\micro\meter} resolution requires a ``tiled" tomography experiment that produces more than 1.7 TB (9K$\times$11K images with 4.5K angles) measurement data. 
Further, the reconstruction of such data generates more than 4.3 TB 3D volumetric image (with 9K$\times$11K$\times$11K voxels)~\cite{Vescovi:il5010}. 
High-performant scalable solvers are needed to reconstruct these large experimental datasets, especially considering that advancements in domain sciences, such as neuroscience, require imaging and reconstruction of many of these samples.

Due to experimental constraints, such as extreme conditions (high radiation dose) or physical limitations (vibrations or drifts during data acquisition), produced data can be far from the ideal and can consist of noisy images with undesired artifacts.  
These imperfect measurements can adversely impact the reconstruction process resulting in unsatisfactory output. 
The ability to mitigate such noise and artifacts 
are important considerations when it comes to choosing between the two broad categories of reconstruction approaches: analytical methods and iterative solvers. 
While analytical methods, such as filtered-backprojection, are typically fast algorithms, they produce sub-optimal reconstructions with imperfect (noisy) measurement data. 
In contrast, iterative tomographic reconstruction solvers enable integration of advanced regularizers and models, and iteratively reach for a solution by solving an optimization problem. 
They also provide an essential framework for incorporating  imperfections into the model 
and therefore, mitigate these artifacts and achieve the desired resolution. 
However, such improved reconstruction comes at the expense of significantly higher computational cost.

Iterative reconstruction 
solvers have so far been used for small/medium scale tomography datasets mostly due to the aforementioned computational costs.
Different parallelization methods have been developed to ease this cost, from naive data-parallel approaches that process different slices of the measurement data (sinograms) in parallel to advanced in-slice and  memory-centric approaches~\cite{bicer2015europar, hidayetouglu2019memxct, wang2019consensus}.
For parallel beam geometry, sinogram-based reconstruction methods exploit data parallelism by reconstructing each 2D slice (sinogram) independently.
After \reb{the} reconstruction is done, all reconstructed slices (tomograms) are gathered to a 3D volume. 
This approach provides reasonable execution time for most smaller datasets, but larger datasets require more  aggressive parallelization. 
In-slice  parallelization techniques improve the  speed of single sinogram reconstruction by distributing parts of a sinogram to several  processes, but introduce synchronization and communication overheads during iterations since the processes that are involved in the processing of the same sinogram need to 
combine their intermediate results.
The MemXCT advanced memory-centric parallelization technique \cite{hidayetouglu2019memxct} mitigates the synchronization and communication overheads by using memoization, and provides efficient sparse matrix representations and communication patterns. 

\begin{figure}[t!]
\begin{subfigure}{.25\textwidth}
  \centering
  \includegraphics[width=.95\textwidth]{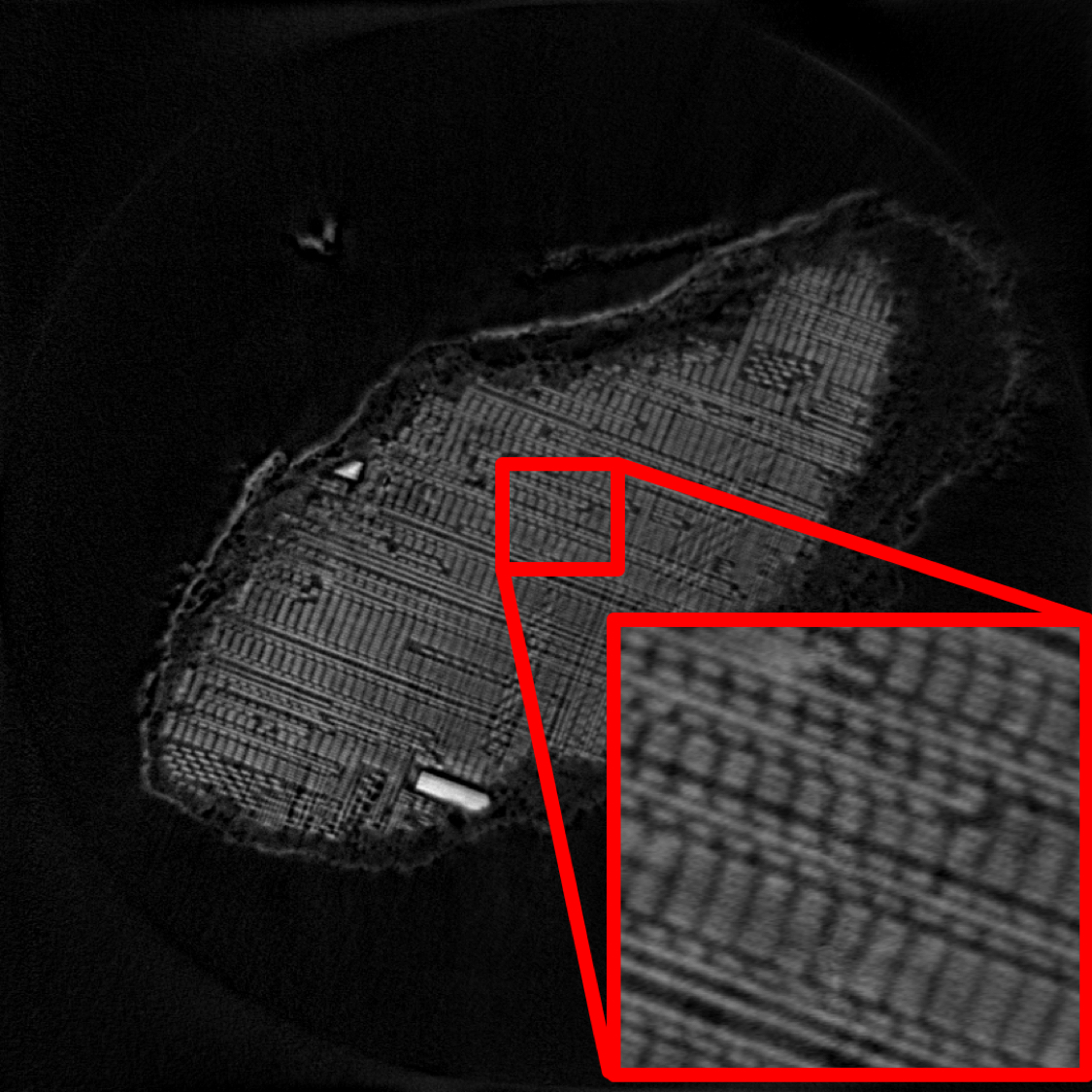}
  \caption{}
  \label{fig:chip}
\end{subfigure}%
\begin{subfigure}{.25\textwidth}
  \centering
  \includegraphics[width=.95\textwidth]{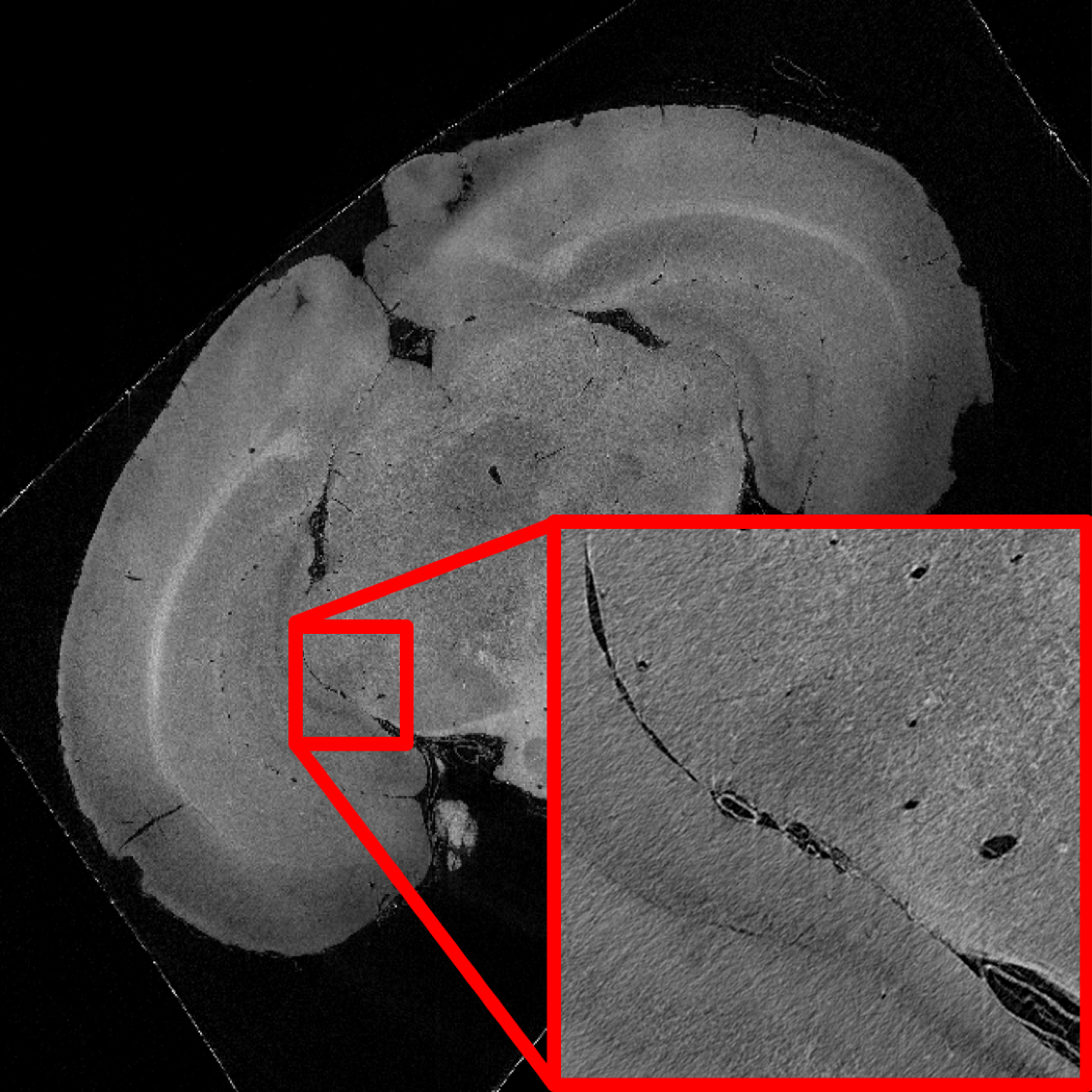}
  \caption{}
  \label{fig:brain}
\end{subfigure}
\vspace{5mm}
\caption{(a) Tilted slice of an IC chip reconstruction and (b) horizontal slice of a mouse brain reconstruction. }
\label{fig:samples}
\end{figure}

While the MemXCT parallelization technique provides significant performance improvement compared to its alternatives, reconstruction of full-sized volumes of extreme-scale samples still requires long processing time.
For example, authors in~\cite{hidayetouglu2019memxct} report that reconstruction of a single mouse brain sinogram (a slice of the measurement data) requires 10 secs using 256K-cores at Theta supercomputer at ANL. 
The full reconstruction of the sample (9K sinograms)
requires more than 25 hours with the whole supercomputer. 
Fig.~\ref{fig:samples}(a) and Fig.~\ref{fig:samples}(b) show reconstructions of a medium-scale integrated circuit (IC) and a large-scale mouse brain tomograms (slices of 3D images), respectively. 
The reconstruction quality is extremely important to distinguish features for these samples (transistors and wires for IC, and blood vessels and myelinated axon tracts for brain sample), therefore iterative solvers are preferred method for reconstruction. Note that a typical science study will likely require the full reconstruction of many samples.

Large-scale GPU resources, such as \reb{the} Summit supercomputer at the Oak Ridge National Laboratory (ORNL) provide opportunities to apply iterative reconstruction algorithms to extremely large tomography datasets. 
However, further optimizations and advanced parallelization techniques must be used to achieve maximum efficiency on such resources. 
In this paper, we introduce novel optimizations for state-of-the-art memory-centric iterative reconstruction approaches to enable reconstruction of extremely large tomography datasets. 
Specifically, we make the following contributions:

\begin{itemize}[leftmargin=*,noitemsep,nolistsep]

\item We introduce improved data partitioning and parallelization techniques that extend 2D MemXCT \emph{data parallelism} with \emph{3D batch parallelism}. Our approach enables optimized SpMM operations that perform common computational kernels on different voxels (fusing), while performing memoization of irregular data accesses at 3D
space (compared to 2D plane in MemXCT).

\item We present a hierarchical communication strategy that exploits multi-GPU node architecture. Our approach leverages asynchronous multi-level data reduction to minimize communication overhead between nodes.

\item We propose mixed-precision implementation that performs computations using single (or double) precision operations, whereas it stores and communicates data in half precision for reduced memory and communication footprint.

\item We provide a comprehensive evaluation of our optimizations and system with four real-world tomography datasets using up to 24K NVIDIA V100 GPUs and demonstrate sustained peta-scale performance.
\reb{In particular, we reconstruct a mouse brain volume with 9K×11K×11K voxels within 3 minutes using the whole Summit supercomputer, reaching 65 PFLOPS throughput (or 34\% of Summit's theoretical peak performance).} 
\end{itemize}

\section{Background}\label{sec:background}
In this section, we briefly explain tomographic data acquisition with synchrotron light sources and the basics of iterative reconstruction process.

\subsection{Tomography Experiments}
\begin{figure}[t!]
    \centering
    \includegraphics[width=0.45\textwidth]{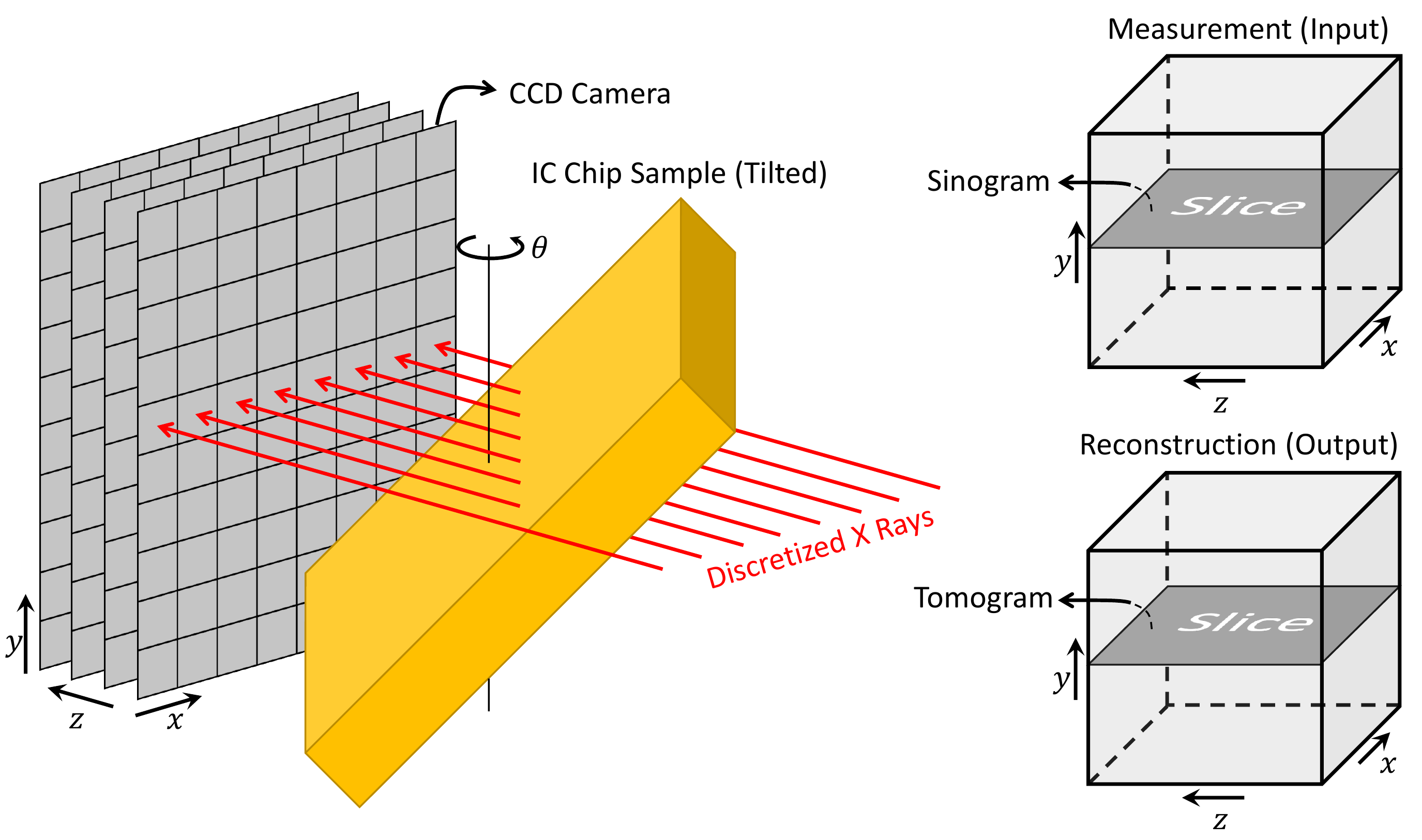}
        \caption{An experimental setup that shows tomographic data acquisition.}
    \label{fig:scanning}
\end{figure}

During a tomography experiment, a sample is placed on top of a rotation stage and exposed to X-ray beams.
As X-ray beams travel through sample, the photons are attenuated by the sample according to Beer-Lambert law~\cite{beer1852bestimmung,lambert1760jh}. 
The attenuated beams are, then, measured at the detector and an X-ray \emph{projection} of the sample is recorded at the detector, as illustrated in Fig.~\ref{fig:scanning}.
This process is repeated for different rotational views of the sample, $\theta$, with the aim of meeting Crowther criterion~\cite{crowther1970reconstruction}. 
Consequently, this process generates a set of projections, $p_\theta$, where, for example, $\theta=\{0\si{\degree}, 1\si{\degree}, 2\si{\degree},\dots, 179\si{\degree}\}$ degrees. 

Iterative tomographic reconstruction aims to solve an optimization problem. The following cost function is often used:
\vspace{-2mm}
\begin{equation}\label{eq:cost}
\hat{x}=\underset{x\in C}{\textrm{argmin}}\{\|y-Ax\|^2_2+R(x)\}
\vspace{0mm}
\end{equation}
Here, $y$ is the measurement data, $x$ represents the estimated versions of the unknown object, $C$ is a constraint on $x$ and \reb{$R(x)$ is a regularizer function.}
$A$ is the system matrix or forward operator that depicts how the X-ray beams intersect the voxels in each rotational view and thus the relationship between $x$ and 
$y$, i.e. the sinograms and the object, respectively. The solution $\hat{x}$ is the version of the estimated object that minimizes the cost function.

A generic optimization solver requires computing the gradients and updating of the object based on those gradients in an iterative fashion. 
First, the \emph{forward operator}, $A$, is applied to previously estimated $x_i$ object, then the result is subtracted from sinograms, $y-Ax_i$, and residual $r_i$ is computed based on a norm, \reb{e.g., Euclidean norm as in equation 1.} 
Next, $r_i$ is \emph{back projected} on $x_i$ to compute the gradients. 
Finally, the estimated object $x_i$ is updated based on the computed gradients and the new object $x_{i+1}$ derived for next iteration.
In this work, we consider parallel beam geometry for experiments at the synchrotrons and that all beams travel perpendicularly to the axis of rotation, 
which enables independent reconstruction of slices  
in 3D object volume from their corresponding sinograms. 
We also perform an optimized version of Siddon's algorithm for accurate forward and back projection operators~\cite{siddon:1985}.


\begin{figure}[!t]
    \centering
    \includegraphics[width=0.45\textwidth]{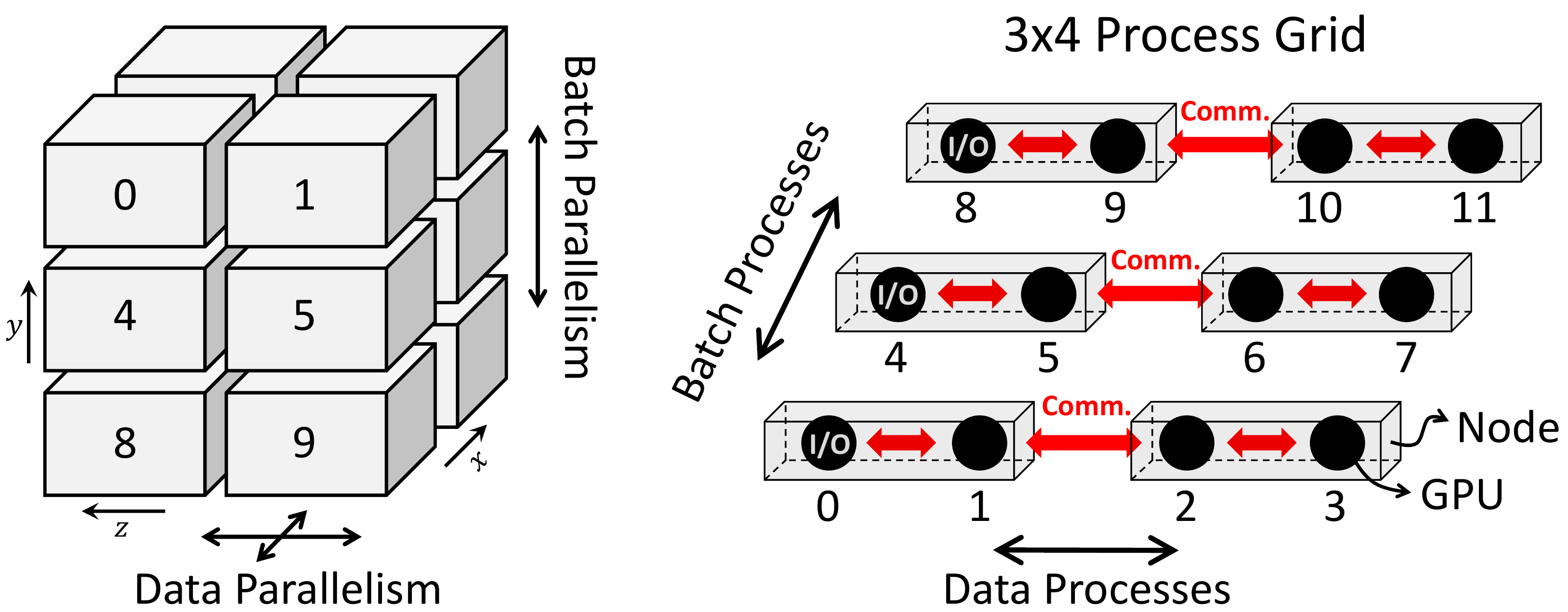}\\
    \small\hspace{-0.4cm}(a)\hspace{4cm}(b)
    \vspace{1mm}
    \caption{\small (a) 3D domain partitioning and (b) assignment on GPUs.} 
    \label{fig:partitioning}
    \vspace{-2mm}
\end{figure}

\subsection{Challenges and Opportunities}
The sparse system matrix $A$ in Eq.~\ref{eq:cost} represents the incident voxels of each ray, and therefore its memory footprint can be large. 
Forward and backprojection operators perform irregular accesses to $A$ and $x_i$ while computing residuals and updating objects. 
These operations introduce significant overhead during the iterations. 
Further, if the $A$ matrix cannot fit into available memory, incident rows and columns need to be computed repetitively before each forward and backprojection operation. Furthermore, if some $A$ and $x$ elements involved in an inner product calculation performed by a process are in the memory of another process, that missing information must be communicated.

Prior work MemXCT addresses communication and irregular data access overheads using multi-level Hilbert ordering to maximize the likelihood that all system matrix $A$ elements involved in an inner product of the sparse matrix-vector multiplication are in the same partition~\cite{hidayetouglu2019memxct}. 
It also avoids repetitive computation of $A$ with memoization while partitioning large memory footprint to many processes. 
However, the MemXCT approach processes each sinogram independently of others, hence the optimizations are performed only within 2D slices.

In this work, we identify and exploit the following optimization opportunities in 3D:
\begin{itemize}[leftmargin=*,noitemsep,nolistsep]
\item Assuming a ray, $u_{i,j}$, where $i$ and $j$ are the location of the ray in the $y$ and $x$ dimensions, respectively, all rays $u_{*,j}$ trace the same voxels in their respective slices for all rotational views.
This property can be used to \emph{fuse} rays along the $y$ dimension so that the A elements can be reused when processing these rays. 
\item MemXCT approaches typically utilize single data type throughout their execution. This can introduce unnecessary overhead during data movement. Mixed precision data types can minimize data transfer volume while meeting precision requirements of computation.
\item Direct communication between reconstruction processes can create redundant overhead. Reducing data between co-located processes can significantly improve data transfer efficiency.
\end{itemize}

\section{Algorithm Design \& Optimizations}\label{sec:design}

In this section, we present five categories of optimizations that exploit the opportunities identified in Section II.

\begin{figure}[t!]
    \centering
     \includegraphics[width=0.48\textwidth]{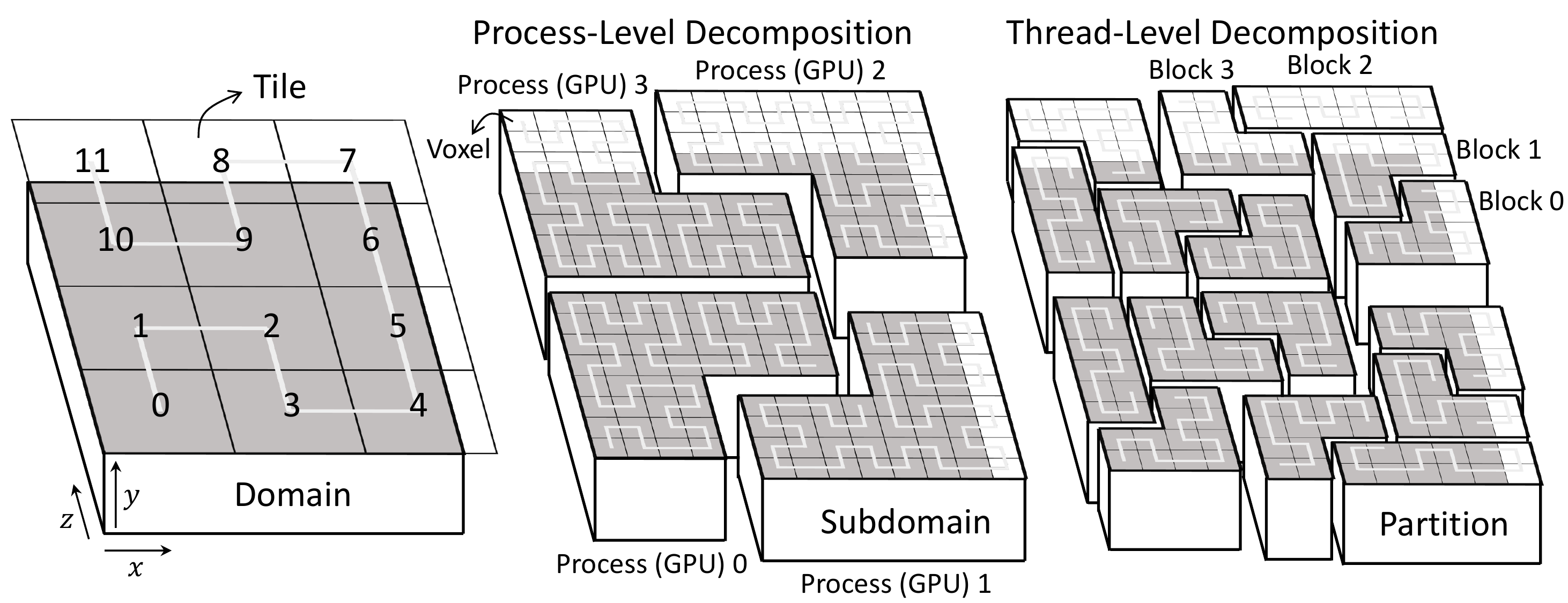}\\
     \small\hspace{0cm} (a) \hspace{2.6cm} (b) \hspace{2.5cm} (c)\\
    \caption{\small 3D Hilbert-ordering domain decomposition at (a) tile (b) process, i.e., GPU, and (c) thread, i.e., block, levels.}\label{fig:domain_decomposition}
\vspace{-2mm}
\end{figure}

\subsection{Partitioning and Parallelization}\label{sec:parallel}

We propose a data partitioning and parallelism arrangement strategy for both input (sinogram) and output (tomogram) data cubes as depicted in Fig.~\ref{fig:partitioning}(a). 
The proposed strategy combines two principal parallelism modalities: \emph{batch parallelism} and \emph{data parallelism}.

Batch parallelism takes advantage of the fact that there is no data dependency between the slices in the $y$ direction. 
Therefore, tomogram and sinogram slices are partitioned equally among \textit{batch processes}\footnote{In the context of this paper, each process corresponds to a GPU.} and reconstructed in an embarrassingly-parallel fashion.
However, because batch parallelism does not involve partitioning of data in the $x$ and $z$ dimensions, the per-process memory footprint of slices may not fit within the GPU memory of batch processes. 

Data parallelism solves this problem by partitioning slices among \textit{data processes} in the $x$-$z$ plane along with their corresponding data structures. 
As a result, per-process memory footprint is reduced so that it can fit within GPU memory. 
Nevertheless, data parallelism comes with a communication overhead because of dependency between data partitions in the $x$ and $z$ dimensions.
The detailed design of the proposed strategy consists of the following components to minimize communication and maximize data reuse.

\subsubsection{Hilbert-Ordering Domain Decomposition}\label{sec:hilbert}

In order to partition a batch of slices in the $x$-$z$ plane while maintaining data locality, we extend the MemXCT 2D Hilbert-ordering domain decomposition to 3D, where it is applied to all tomogram and sinogram slices in the $y$ direction. 
Fig.~\ref{fig:partitioning} and Fig.~\ref{fig:domain_decomposition} depict decomposition  at the process and thread levels: First, both tomogram and sinogram domains are tiled into square patches and ordered with pseudo-Hilbert ordering.
These patches are partitioned equally among processes, where each partition corresponds to a \textit{subdomain} as shown in Fig.~\ref{fig:domain_decomposition}(b). Each subdomain is processed by a single GPU.

Fig.~\ref{fig:partitioning}(a) shows that a domain is partitioned into \reb{12} subdomains: \reb{two} in the $x$ direction, \reb{two} in the $z$ direction, and \reb{three} in the $y$ direction. Fig.~\ref{fig:partitioning}(b) depicts assignment of these 12 subdomains to a GPU grid with 12 processes.
The GPU grid consists of six nodes and each node contains two GPUs. 
Since communications within node have a higher bandwidth than that of between nodes, data processes processing adjacent subdomains are placed in the same node whenever possible.

Then each subdomain is partitioned among GPU thread blocks as shown in Fig.~\ref{fig:domain_decomposition}(c).
Data connectivity and locality of partitions provided by the Hilbert-ordering domain decomposition are essential for optimized SpMM design (Sec.~\ref{sec:computation}) and hierarchical communications (Sec.~\ref{sec:communication}) -- two of the major contributions of this paper.

\subsubsection{Batch Processing Pipeline}\label{sec:pipeline}
To overlap MPI communication and GPU computation, we partition each batch into smaller \textit{I/O batches} that are processed sequentially, i.e., one I/O batch is reconstructed at a time. As discussed in Sec.~\ref{sec:computation}, optimzed SpMM fuses multiple slices in a I/O batch into \textit{minibatches}.
Processing of these minibatches are pipelined by overlapping MPI communications and GPU computations as explained in Sec.~\ref{sec:overlapping}. 
To achieve sufficient overlapping, there needs to be at least a few minibatches in an I/O batch.

\subsubsection{Optimal Partitioning Strategy}\label{sec:optimal}
In order to minimize the communication overhead of data parallelism, it is better to minimize partitioning of the 3D data cube in the $x$-$z$ dimension; only until per-process memory footprint fits into GPU memory. 
Then batch partitioning should take over in the $y$ dimension with no additional overhead. 
The level of batch parallelism is limited by the total number of slices in the $y$ direction and the need to fuse slices so that \reb{the sparse matrix $\boldsymbol{A}$ can be reused from register by the optimized SpMM.}

\begin{table}[h!]

\vspace{2mm}
\captionsetup{font=small}
\vspace{1mm}
\caption{Computational Complexity}
\label{tab:complexity}
\vspace{1mm}
\centering
\hspace{1cm}\begin{tabular}{r|c|c}
 & \textbf{Per Process} & \textbf{Total} \\\hline
\textbf{Comput.} & $MN^2/P_bP_d+MN/P_b\sqrt{P_d}$ & $MN^2+MN\sqrt{P_d}$\\
\textbf{Memory} & $N^2/P_d+N/\sqrt{P_d}$ & $N^2P_b+NP_b\sqrt{P_d}$\\
\textbf{Comm.}* & $MN/P_b\sqrt{P_d}$ & $MN\sqrt{P_d}$ \\
\end{tabular}

\vspace{1mm}
\begin{tablenotes}
    \captionsetup{font=footnotesize}
    \item $M$ row channels, $N$ column channels, $P_b$ batch processes, $P_d$ data processes.
    \item *Latency and contention terms are omitted.
\end{tablenotes}
\vspace{-1mm}
\end{table}

\subsubsection{Computational Complexity}\label{sec:complexity}
The computational cost of a projection depends on size of the 2D detector grid depicted in Fig.~\ref{fig:scanning}. Here, $M$ and $N$ are the number of row and column channels in the detector grid. Since each \reb{discretized ray measured by a detector propagates} through  $\mathcal{O}(N)$ voxels, each projection takes $\mathcal{O}(MN^2)$ time. 
Parallel rays have the same trajectory in all slices, and therefore it is sufficient to store a single sparse matrix with $\mathcal{O}(N^2)$ nonzeroes and reuse it from memory for all $M$ slices. 
\reb{The total memory footprint is increased by a factor of batch processes ($P_b$) because of the sparse matrix duplication.} 
On the other hand, data parallelization partitions the sparse matrix into $P_d$ data processes, where each process computes a partial projection. 
The geometrical shapes of the partial data is shown in \reb{Fig.~\ref{fig:communication_direct}(b)}. 
The partial data of a processes scales with $MN/\sqrt{P_d}$ because the cross-section of each subdomain on the detector halves only when $P_d$ is quadrupled. 
To obtain the total projection, the partial data is communicated among all data processes to be reduced at the receiving data processes. 
This additional communication and reduction is the overhead of data parallelism. 
Table~\ref{tab:complexity} summarizes per-process and total computation, memory, and communication complexities of the proposed 3D partitioning.

\subsection{XCT-Optimized SpMM Design}\label{sec:computation}

By considering memory access patterns specific to XCT, we optimize the proposed \reb{mixed-precision SpMM} kernel via a 3D input/output buffering algorithm. 
This subsection discusses these optimizations and explains our CUDA implementation given in Listing~\ref{listing:optimized_spmm}: The same implementation is used for both projection and backprojection.

\begin{figure}[t!]
    \centering
     \includegraphics[width=1\columnwidth]{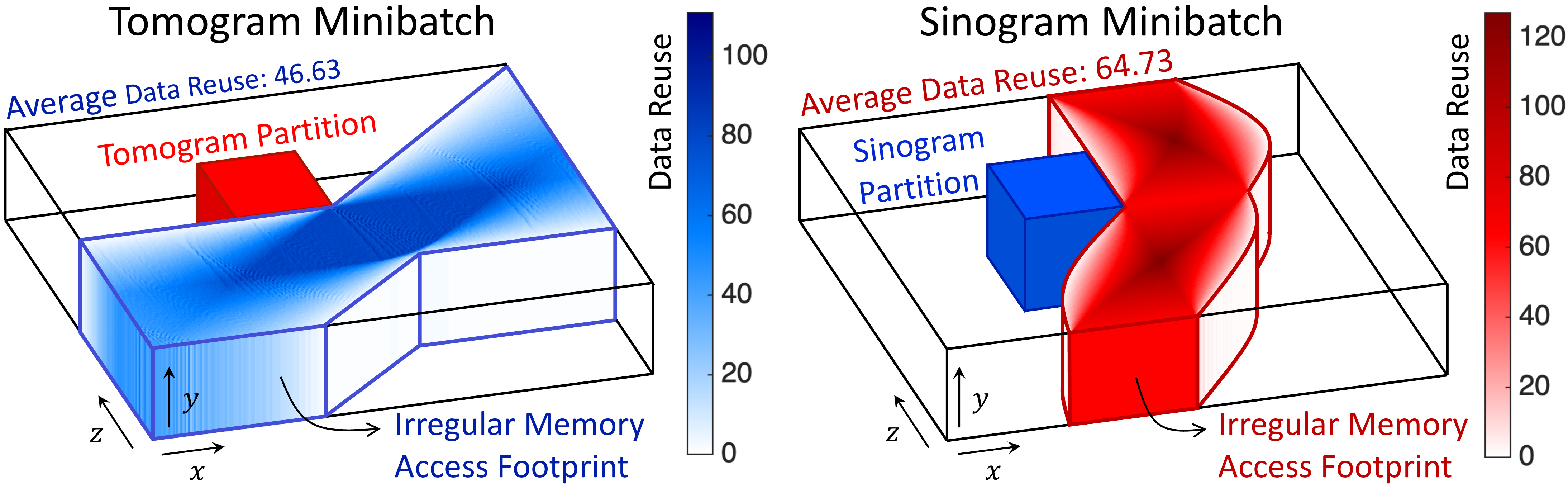}\\\vspace{-1mm}
    \small(a)\hspace{3.8cm}(b)\\\vspace{1mm}
    \includegraphics[width=1\columnwidth]{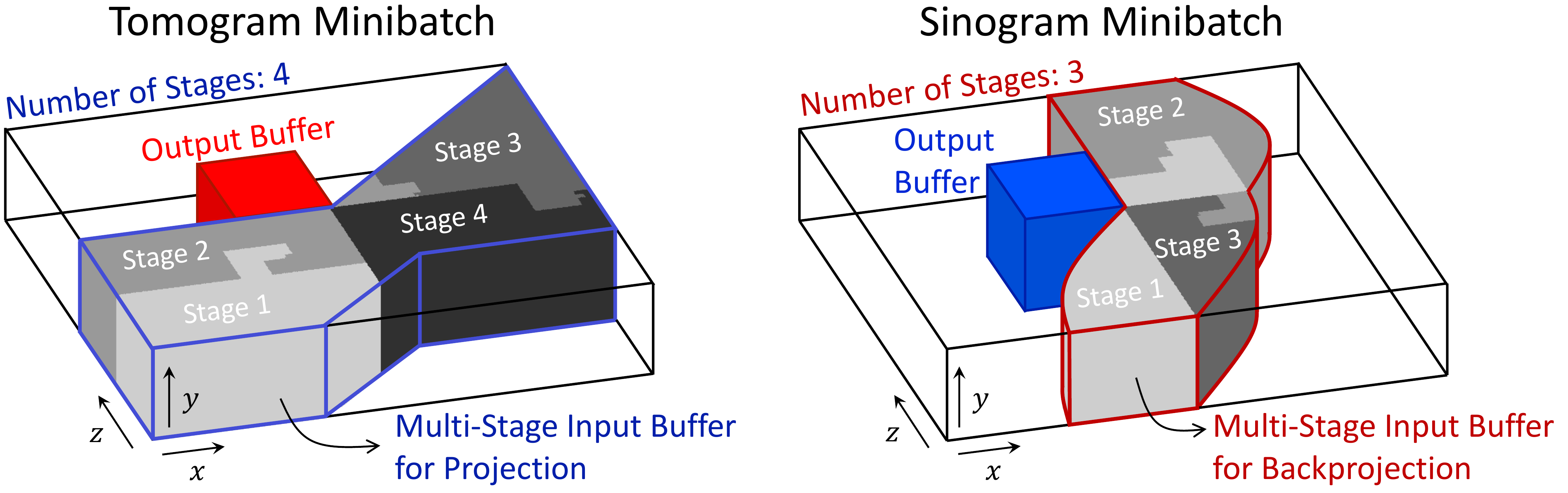}\\\vspace{-1mm}
    \small(c)\hspace{3.8cm}(d)\\\vspace{1mm}
    \caption{ (a, b) Tomogram and sinogram partitions and respective data access footprints, and (c, d) multi-stage buffer shapes.}\label{fig:access_pattern}
\end{figure}

\subsubsection{3D Input Buffering}\label{sec:input_buffering}
A naive SpMV implementation suffers not only from irregularity but also redundant accesses to tomogram and sinogram data because each voxel is (irregularly) accessed more than once by different threads. As a remedy, our optimized kernel stages data accesses in the \reb{GPU} shared memory via 3D input buffers.  
Fig.~\ref{fig:access_pattern} illustrates the basic idea of our optimization by depicting memory access footprints for a 256$\times$256$\times$50 minibatch; (a) shows tomogram voxels accessed by a sinogram \reb{partition} and (b) shows sinogram voxels accessed by a tomogram partition, where each partition is computed by a single thread block. 
In 3D input buffering, each thread block loads the highlighted input data (Fig.~\ref{fig:access_pattern}: (a) and (b)) to shared memory once, and reuses it (irregularly) from shared memory. As a result, memory accesses are reduced by the factor of data reuse shown in the figure, \reb{i.e., darker shade represents higher data reuse from shared memory}. With these reuses, the GPU performance becomes limited by the memory bandwidth consumed when reading $\boldsymbol{A}$ and $\boldsymbol{A}{}^T$ from memory (see the roofline analysis in Sec.~\ref{sec:roofline}). We overcome this memory-bandwidth bottleneck by reusing $\boldsymbol{A}$ elements from registers as we discuss next. 

\subsubsection{Register Reuse}
Even with input buffering, the optimized  SpMV utilizes only less than two percent of the GPU's theoretical peak compute throughput because of its low arithmetic intensity (FLOPS/byte). This is because each fused multiply add (FMA) requires two data elements from memory: (1) shared-memory \textit{index} of the visited voxel by X-ray and (2) \reb{intersection} \textit{length} of the ray going through the voxel. To increase the arithmetic intensity, we propose to reuse index and value data from registers.
That is, many SpMVs in a minibatch are fused as $\boldsymbol{A}\boldsymbol{X}=\boldsymbol{B}$, where each column of $\boldsymbol{X}$ and $\boldsymbol{B}$ corresponds to a slice of tomogram and sinogram values in the minibatch, respectively. As a result, the arithmetic intensity of the operation increases by the fusing factor, a.k.a., minibatch size. However, the fusing factor cannot be arbitrarily large because fusing imposes register pressure as discussed next. 

\subsubsection{3D Output Buffering}\label{sec:register_pressure}
To provide data reuse from register, each thread loads one index and one length at a time and then reuses this pair from register for all corresponding output voxels in the \reb{$y$} direction of the minibatch. To accumulate the partial sums, each thread allocates an output buffer (\texttt{acc} in Line 10 of Listing~\ref{listing:optimized_spmm}). Line 28 shows how a thread reuses the index and value pairs to access its corresponding data in the 3D input buffer. 
\reb{However, since each streaming multiprocessor (\reb{SM}) has a limited number of registers that are used by all threads, enlarging the minibatch size, i.e., \texttt{FFACTOR}, can increase the number of registers required by each thread and hence elevate register pressure on the SMs.} 
When threads collectively use more registers than available in \reb{SM}, register contents are spilled to slower memory and hamper GPU performance. 
Results show that GPUs are able to obtain 34\% of their theoretical compute throughput by carefully tuning the minibatch size as shown in Sec.~\ref{sec:computation} 

\begin{lstlisting}[language=C++, caption={\small Optimized SpMM Mixed-Precision Kernel
\label{listing:optimized_spmm}},basicstyle=\notsotiny\ttfamily,tabsize=1,literate={\ \ }{{\ }}1,numbers=left,xleftmargin=6ex, frame=t!,morekeywords={half, matrix}]
//MATRIX STRUCTURE
struct matrix{ unsigned short ind; half len; };
//PROJECTION KERNEL
__global__ void kernel_project(half *y, half *x, matrix *mat, 
                                                    int *displ, int numrow, int numcol,
                                                    int *buffdispl, int *buffmap,
                                                    int *mapdispl, int *mapnz,
                                                    int buffsize ){
    extern __shared__ half shared[];
    <@\textcolor{codered}{\textbf{float} acc[FFACTOR] = {0.0};}@>
    int wind = threadIdx.x%WARPSIZE;
    for(int buff = buffdispl[blockIdx.x]; buff < 
                                                                buffdispl[blockIdx.x+1]; buff++){
        int mapoffset = mapdispl[buff];     
        for(int i = threadIdx.x; i < mapnz[buff]; i += blockDim.x){
            int ind = buffmap[mapoffset+i];
            <@\textcolor{codeblue}{\#pragma unroll}@>
            for(int f = 0; f < FFACTOR; f++)
                shared[f*buffsize+i] = x[f*numcol+ind];
        }
        __syncthreads();
        int warp = (buff*blockDim.x+threadIdx.x)/WARPSIZE;
        for(int n = displ[warp]; n < displ[warp+1]; n++){
            <@\textcolor{codegreen}{\textbf{matrix} mat = indval[n*WARPSIZE+wind];}@>
            <@\textcolor{codered}{\textbf{float} len = \_\_half2float(mat.len);}@>
            <@\textcolor{codeblue}{\#pragma unroll}@>
            for(int f = 0; f < FFACTOR; f++)
                <@\textcolor{codered}{acc[f] += \_\_half2float(shared[f*buffsize+mat.ind])*len;}@>
        }
        __syncthreads();
    }
    int row = blockIdx.x*blockDim.x+threadIdx.x;
    if(row < numrow)
        <@\textcolor{codeblue}{\#pragma unroll}@>
        for(int f = 0; f < FFACTOR; f++)
            <@\textcolor{codered}{y[f*numrow+row] = \_\_float2half(acc[f]);}@>
}
\end{lstlisting}

\subsubsection{Multi-Stage 3D Buffering}
Each \reb{SM} has a limited capacity of shared memory (96 KB for V100 GPUs), which is not enough for large memory access footprint of a 3D partition. Therefore we provide access to input data in multiple stages. Fig.~\ref{fig:access_pattern}: (c) and (d) show four-stage and three-stage bufferings for projection and backprojection, respectively. Each stage loads 96 KB data whose shapes are governed by Hilbert ordering as described in Sec.~\ref{sec:hilbert}. Line 12 in Listing \ref{listing:optimized_spmm} shows the iterations over stages, where only a portion of the 3D input buffer is loaded \reb{by mapping \texttt{buffmap}}, and only a partial value of the 3D output buffer is computed in each stage. Because each staging has a synchronization overhead (seen in Lines 21 and 30), fewer number of stages is desirable. 
The 3D buffering strategy increases the number of stages since memory footprint of input buffers grow with increasing minibatch size (in the $y$ direction). We mitigate increasing memory footprint using mixed precision implementation as explained in the following subsection.

\subsection{Mixed-Precision Implementation}\label{sec:mix-precision}
Mixed-precision implementation stores and communicates data in half precision (16-bit) and performs all arithmetic operations with single precision (32-bit) through in-core data conversions as in highlighted in Listing~\ref{listing:optimized_spmm} (red). 
This approach reduces memory and communication footprint \reb{as well as} provides \reb{a} higher GPU throughput by moving data in half-precision while maintaining numerical accuracy by performing FMAs in single-precision. 
Several issues need to be addressed while using mixed precision implementation:
(1) Half-precision has a lower range and quantization, and therefore it can suffer from overflow and underflow. 
(2) When reading sparse matrix $\boldsymbol{A}$ from memory, each warp of 32 threads accesses only 64 \reb{bytes} of data at a time, which only utilizes half of the 128-byte GPU cache-line and therefore results in sub-optimal cache utilization.
We address these issues with \emph{adaptive normalization} and \emph{data packing}, respectively.

\begin{figure*}[t!]
    \centering
    \includegraphics[width=1\textwidth]{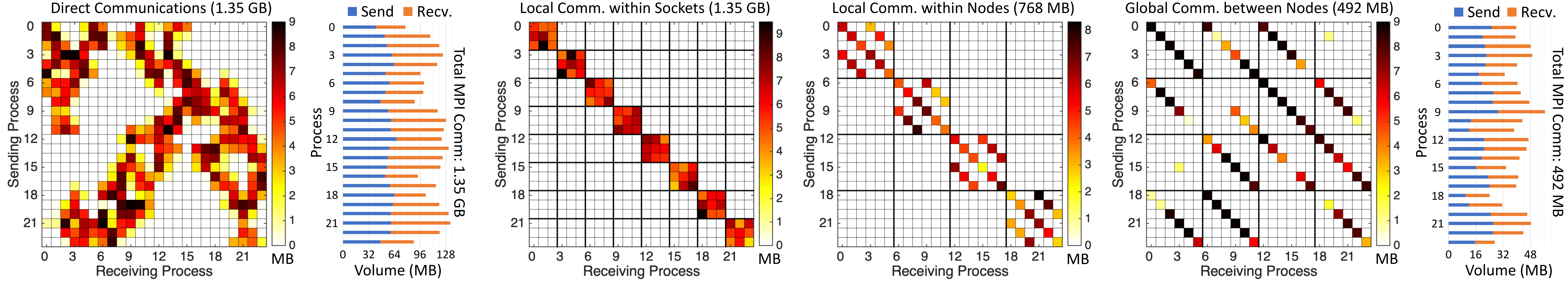}\\
    \small \hspace{-1.8cm} (a) \hspace{5cm} (b) \hspace{3cm} (c) \hspace{3cm} (d)\\
    \vspace{1mm}
    \caption{(a) Direct communication and load balancing. Three-level hierarchical communications: (b) Socket-level communication; (c) Node-level communication; (d) Global communication and load balancing.}
    \vspace{-1mm}\label{fig:communication_hierarchical}
\end{figure*}

\begin{figure}[t!]
    \centering
     \hspace{-1mm}\includegraphics[width=\columnwidth]{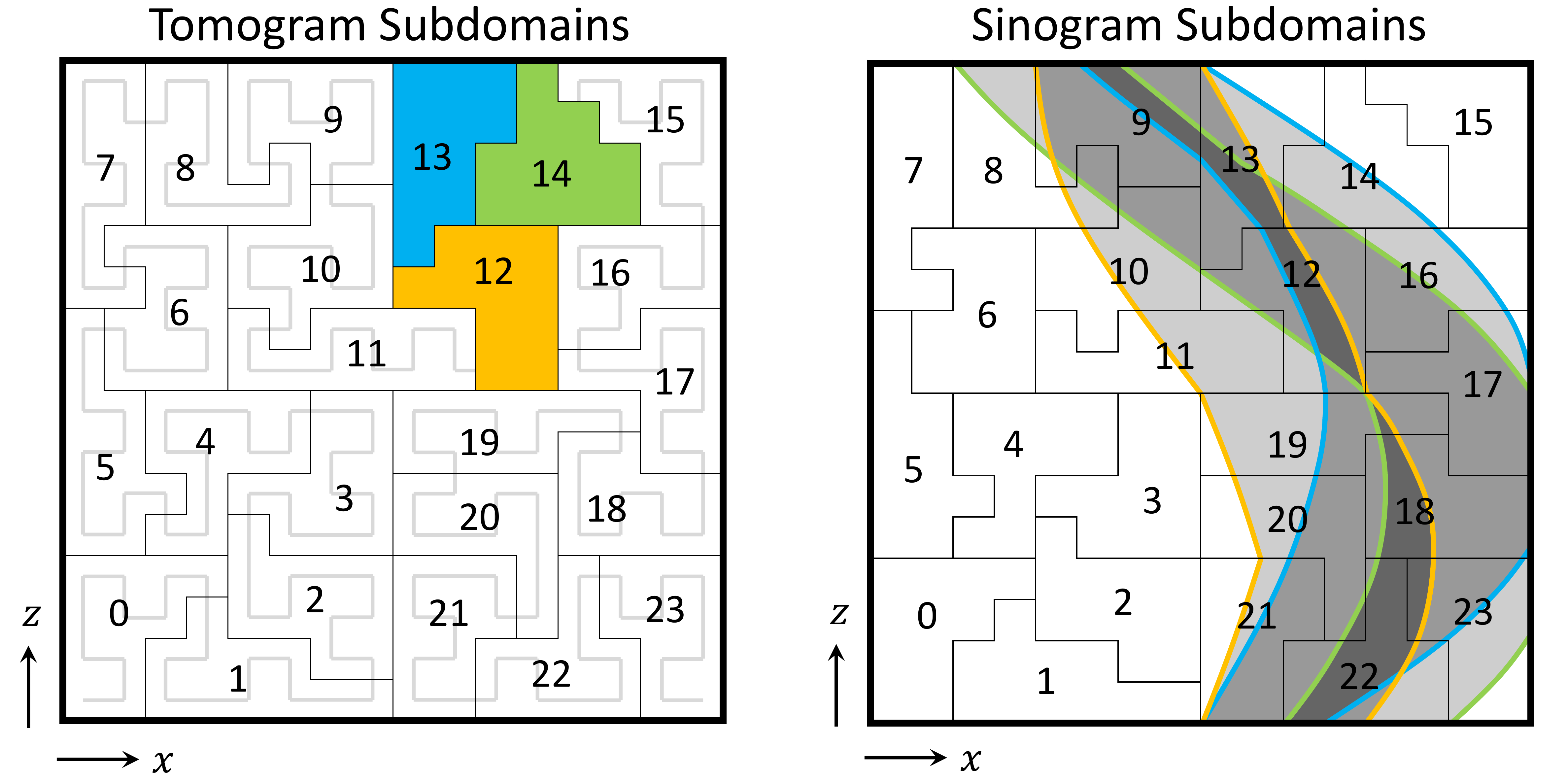}\\
    \small\vspace{-1mm}\hspace{0mm}(a)\hspace{3.8cm}(b)\\\vspace{2mm}

    \caption{ Tomogram (a) and sinogram (b) subdomains. Partial data footprint of tomogram subdomains 12--14 are shown in (b). }\label{fig:communication_direct}
    \vspace{2mm}
\end{figure}

\subsubsection{Adaptive Normalization} 
We avoid the half-precision underflows by \reb{artificially} increasing the voxel size in the tomogram.
Since the nonzero sparse matrix values represent the travel lengths of incident X-rays per voxel, increasing the voxel sizes results in larger half-precision values and hence avoids underflows. 
On the other hand, the overflows are handled by maintaining input and output data in double or single precision before performing kernel operations. We accomplish this by normalization and denormalization of input and output data before and after type castings, respectively. 
The (de)normalization factor is adaptively changed \reb{in each iteration} with respect to the max-norm of the evolving input vector to prevent overflows while minimizing underflows.


\subsubsection{Data Packing}
The underutilization of GPU cache line not only wastes memory bandwidth but also introduces latency. As a remedy, we pack both index and length data elements into a single structure of four bytes. 
This structure is shown in Line 2 of Listing~\ref{listing:optimized_spmm}: It consists of an unsigned two-byte integer that represents the voxel index in shared-memory, and a two-byte half-precision floating point number representing the length. As a result, each warp of 32 threads utilizes the full 128-byte cache line.


\subsection{Hierarchical Communications}\label{sec:communication}
Data must be partitioned when solving large problems in order to fit per-process memory footprint within available GPU memory. However, data partitioning requires communication and reduction of partial data, as elaborated in Sec.~\ref{sec:complexity}. Consequently, image reconstruction time is dominated by communication overhead for large problems. As a solution, we propose a novel hierarchical communication with  partial data reduction, another major contribution of this paper. This strategy is designed for multi-GPU \reb{node} architectures where high-bandwidth connections between \textit{peer} GPUs are exploited. 
This section describes direct communication of partial data (the baseline), the main idea behind local communication and reduction of partial data, our communication hierarchy, and its efficient implementation.

\subsubsection{Direct Communications}\label{sec:baseline}
Fig.~\ref{fig:communication_direct} shows partitioning of (a) tomogram data and (b) sinogram data into 24 subdomains. The subdomain shapes are governed by Hilbert-ordering domain decomposition (Sec.~\ref{sec:hilbert}). In projection, each process computes only a partial sinogram data which needs to be communicated and reduced to find total sinogram data. As an example, Fig.~\ref{fig:communication_direct}(b) shows the partial data footprints of processes 12--14 on sinogram 
subdomains, e.g., process 12 sends its corresponding partial data to processes 8--13 and 18--23. 
The resulting communication matrix is shown in Fig.~\ref{fig:communication_hierarchical}(a). Communication volume between two processes depends on the amount of overlapped area between sender's partial data footprint and receiver's sinogram subdomain. As a result, the communication is sparse and irregular. Following the communication, receiving processes \textit{reduce} (sum) the overlapping partial data coming from sender processes, which completes projection operation. This description is also valid for backprojection as it is a transpose of projection. 

Fig.~\ref{fig:communication_hierarchical}(a) also shows communication volume of each process and total amount of communicated partial data. Large problems communicate a large portion of data through slow interconnect between nodes, which constitutes a dominating bottleneck. We 
relieve this bottleneck by reducing partial data locally \reb{within nodes and communicating the reduced data between} nodes, as we shall discuss next.

\subsubsection{Local Reduction of Partial Data}
The proposed hierarchical communication exploits high-bandwidth connections between GPUs within nodes by a \textit{local} communicator and reduction of the overlapping partial data among sender processes. To explain, we consider processes 12--14 (Fig.~\ref{fig:communication_direct}(a)) located in the same node. Overlapping portions of their partial data (Fig.~\ref{fig:communication_direct}(b)) are communicated and reduced within the node, rather than communicating the original partial data among nodes directly. Then, the reduced partial data within the node is sent to receivers by a \textit{global} communication, where they are further reduced to find total sinogram values. The locality of subdomains provided by Hilbert ordering is essential in this hierarchical communication because spatially-local subdomains yield more overlapping data and more local partial reductions. Similarly, fatter nodes with more highly-connected GPUs yield more local reduction of partial data.

\subsubsection{The Three-Level Hierarchy}
This paper considers Summit's node architecture, 
where each node has two CPU \textit{sockets}. Each socket is connected to three GPUs that are densely-connected with high-bandwidth interconnect. 
The CPU sockets within a node are connected with a slower data bus. 
Each node is connected to an even slower infiniband network. 
Although, we explain our hierarchical communication and reduction in the context of the Summit architecture, the method is general and applicable to other node architectures with different number of sockets and GPUs. 

Since the data bus between sockets is slow, we do not perform local reduction directly among six GPUs within a node. 
Instead, we first perform a socket-level communication and reduction among three GPUs within a socket. Then, an additional node-level communication and reduction among six GPUs within a node.
Finally, a global communication and reduction among all GPUs between nodes is performed.

Fig.~\ref{fig:communication_hierarchical}(b) presents the socket-level communication matrix as a block-diagonal structure because each GPU talks only to three GPUs (including itself) in the same socket. The socket-level reduction reduces the original 1.35 GB partial data down to 768 MB (43\% reduction). Then, the reduced partial data is further reduced within nodes among six GPUs. Fig.~\ref{fig:communication_hierarchical}(c) shows the intra-node local communication matrix. After the node-level communication, the partial data is further reduced to 492 MB (36\% additional reduction). Finally, GPUs send reduced partial data only among nodes as shown in Fig.~\ref{fig:communication_hierarchical}(d). As a result, the total data communicated among slowly-connected nodes is reduced by 64\% compared to direct communication.

\subsubsection{Efficient Implementation}
To implement the proposed three-level hierarchical communication, we leverage \texttt{MPI\_Comm\_split} to define local communicators within sockets and nodes, and a global communicator among nodes, respectively. To prevent data from being staged through the CPU during local communications, we use CUDA inter-process communication (IPC) capability, i.e., communicating data directly from GPU to GPU bypassing CPU. Furthermore, local communications are overlapped using CUDA streams. Then local reductions are performed on GPUs. Global communications are implemented with MPI with CPU staging through pinned buffers. 
We employ non-blocking \texttt{MPI\_Issend} and \texttt{MPI\_Irecv} for overlapping global communications. 
The partial data is then moved back to the GPUs at the receiving processes for global reductions.

\begin{figure}[h!]
\vspace{-1mm}
    \centering
     \vspace{1mm}\hspace{-1mm}\includegraphics[width=1\columnwidth]{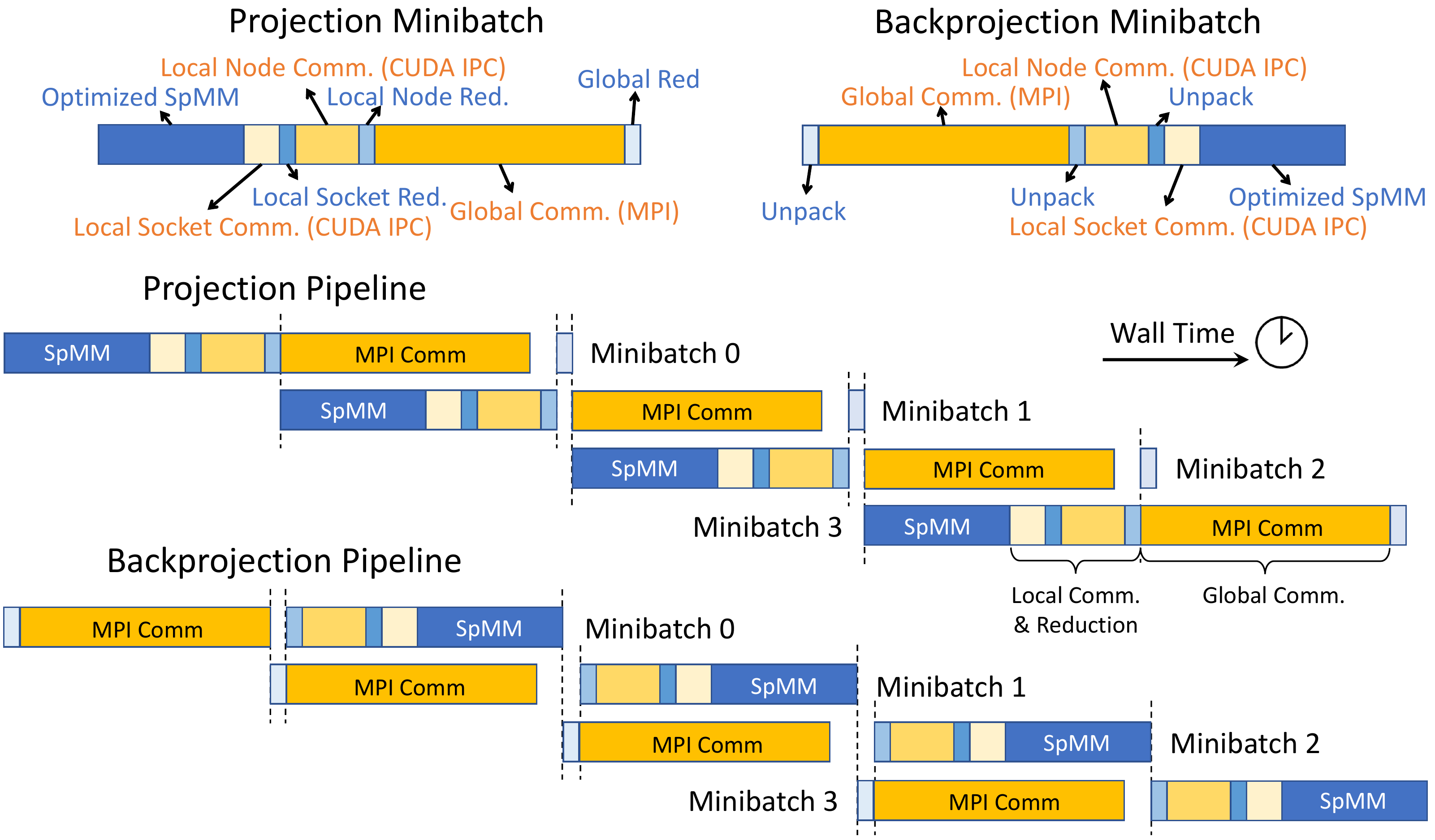}\\
     \vspace{1mm}
    \caption{Overlapping four minibatches in the batch processing pipeline. }\label{fig:overlapping}
    
\vspace{-1mm}
\end{figure}

\subsection{Communication Overlapping}\label{sec:overlapping}
Even when communication time is reduced by applying hierarchical communication strategy, iterative solution time remains bounded by MPI communications for large problems (see Fig.~\ref{fig:communication}). To further alleviate this communication bottleneck, we propose an overlapping strategy that exploits multiple slices in a reconstruction batch which is only possible in 3D image reconstruction. That is, global (MPI) communication of a minibatch is overlapped with local operations of another minibatch involving the optimized SpMM kernel, local reduction kernels, and local (socket-level and node-level) communications. Fig.~\ref{fig:overlapping} depicts overlapping of four minibatches during projection and backprojection operations. Projection overlaps global MPI communication of a minibatch with local operations of the next minibatch. On the other hand, backprojection overlaps local operations with next minibatch's MPI communication. This strategy is the most effective when kernel time and communication time are comparable.

\section{Experimental Results}\label{sec:results}
This section introduces an extensive evaluation of our approaches. 
We first evaluate the performance of our optimizations and designs on Summit supercomputer using four real-world experimental datasets. 
Then, we investigate strong and weak scaling properties of our system. 
Finally, we compare convergence of our optimizations with four different levels of precision.

\subsection{Experimental Setup}

\subsubsection{Summit Supercomputer}
All experiments are conducted on Summit supercomputer at computing facility of ORNL. Summit consists of 4,608 IBM AC922 nodes with two POWER9 CPUs and six NVIDIA V100 GPUs per node. Nodes are connected via dual-raid fat-tree network. 
Each CPU in the node is densely-connected to three GPUs with Nvidia NVlinks, where each link has 50 GB/s one-way and 100 GB/s bidirectional bandwidth. We refer to each CPU and its corresponding densely-connected GPUs as a \textit{socket} in this paper. Each node consists of two sockets. Sockets are connected via an X bus with 64 GB/s bidirectional theoretical bandwidth. Each CPU has 22 cores and 250 GB memory and each GPU has 80 \reb{SM} and 16 GB memory. 

\begin{figure*}[t!] 
    \centering
    \includegraphics[width=1\textwidth]{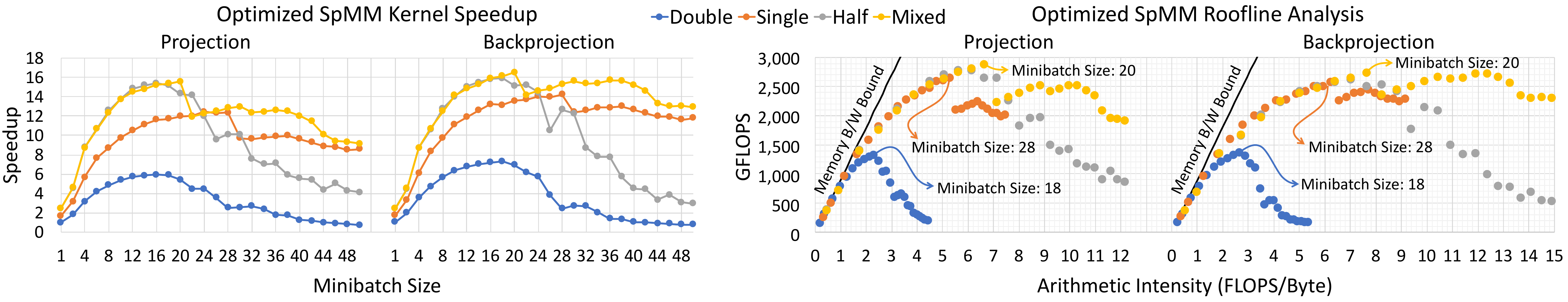}\\
    \small\hspace{0cm}(a)\hspace{9cm}(b)\\
    \vspace{1mm}
    \caption{(a) Speedup and (b) per-GPU performance of XCT-Optimized SpMM kernel as a function of the slice fusing factor, i.e. minibatch size. Speedup is based on
    double-precision projection with no optimization (fusing factor 1).}
    \label{fig:speedup}
\end{figure*}

\begin{table}[h!]
\vspace{3mm}
\centering
\caption{Datasets and Memory Footprints}
\label{tab:dataset}
\resizebox{\columnwidth}{!}{
\begin{tabular}{cccc}
  & \textbf{Measurement Data}  & \textbf{I/O Data} & \textbf{Memory} \\
  \textbf{Sample} & \textbf{Cube} ($K$$\times$$M$$\times$$N$) & \textbf{Footprint} & \textbf{Footprint} \\\hline
Shale Rock & \num{1501}$\times$\num{1792}$\times$\num{2048} & 52.1 GB & 120 GB\\
IC Chip & \num{1210}$\times$\num{1024}$\times$\num{2448}  & 36.7 GB & 139 GB\\
Activated Charcoal & \num{4500}$\times$\num{4198}$\times$\num{6613} & 1.23 TB & 2.82 TB\\
Mouse Brain & \num{4501}$\times$\num{9209}$\times$\num{11283} & 6.56 TB & 10.9 TB
\end{tabular}}%
\end{table}

\subsubsection{Datasets Used for Experiments}
Table~\ref{tab:dataset} characterizes the four datasets used in our evaluation\footnote{In the rest of the paper, we refer to these datasets as {\tt Shale}, {\tt Chip}, {\tt Charcoal}, and {\tt Brain}, according to their order in Table~\ref{tab:dataset}.}.  Dimensions are given in $K$$\times$$M$$\times$$N$, where $K$ denotes the number of projections and $M$ and $N$ denote the number of vertical and horizontal channels in the 2D detector grid, respectively. All measurements follow parallel beam scan geometry as described in Sec.~\ref{sec:background}. The corresponding I/O data and memory footprints are given for all datasets for single precision. All measurements are obtained by experiments at APS, ANL. The {\tt Shale} and {\tt Charcoal} datasets are open, while {\tt Chip} and {\tt Mouse} are proprietary~\cite{tomobank, du2018x, Vescovi:il5010}. \reb{Even though \texttt{IC Chip} and \texttt{Shale} have similar dimensions, we prefer to provide computational performance for the freely available \texttt{Shale} for benchmarking purposes. Nevertheless, we use \texttt{IC Chip} for the convergence results (Sec.~\ref{sec:convergence}) because it is a numerically challenging case with contaminating noise.}



\begin{table}[t!]
\vspace{5mm}
\centering
\caption{Overall Reconstruction Speedup}
\vspace{1mm}
\label{tab:overall}
\resizebox{\columnwidth}{!}{%
\begin{tabular}{cc|ccc|ccc}
\multicolumn{1}{l}{} & \multicolumn{1}{l}{} & \multicolumn{3}{c}{\textbf{\texttt{Shale} on Four Nodes}} & \multicolumn{3}{c}{\textbf{\texttt{Charcoal} on 128 Nodes}}\\
 & \textbf{Prec.} & \textbf{Part.*} & \textbf{Recon.} & \textbf{Speed.} & \textbf{Part.*} & \textbf{Recon.} & \textbf{Speed.} \\\cline{2-8}
\multirow{3}{*}{\textbf{Part. Opt.}} & Double & \textbf{1}$\times$(\textbf{4}$\times$6) & 979 s & 1$\times$ & \textbf{1}$\times$(\textbf{128}$\times$6) & 78.4 m & 1$\times$ \\
 & Single & \textbf{2}$\times$(\textbf{2}$\times$6) & 405 s & 2.42$\times$ & \textbf{2}$\times$(\textbf{64}$\times$6) & 31.3 m & 2.51$\times$ \\
 & Mixed & \textbf{4}$\times$(\textbf{1}$\times$6) & 215 s & 4.56$\times$ & \textbf{4}$\times$(\textbf{32}$\times$6) & 15.1 m & 5.20$\times$ \\\hline
\multirow{3}{*}{\shortstack{Part.$+$\\\textbf{Kernel Opt.}}} & Double & \textbf{1}$\times$(\textbf{4}$\times$6) & 513 s & 1.91$\times$ & \textbf{1}$\times$(\textbf{128}$\times$6) & 58.4 m & 1.34$\times$ \\
 & Single & \textbf{2}$\times$(\textbf{2}$\times$6) & 134 s & 7.30$\times$ & \textbf{2}$\times$(\textbf{64}$\times$6) & 20.4 m & 3.85$\times$ \\
 & Mixed & \textbf{4}$\times$(\textbf{1}$\times$6) & 51.1 s & 19.2$\times$ & \textbf{4}$\times$(\textbf{32}$\times$6) & 8.0 m & 9.78$\times$ \\\hline
\multirow{3}{*}{\shortstack{Part.$+$\\Kernel$+$\\\textbf{Comm. Opt.}}} & Double & \textbf{1}$\times$(\textbf{4}$\times$6) & 218 s & 4.49$\times$ & \textbf{1}$\times$(\textbf{128}$\times$6) & 27.0 m & 3.00$\times$ \\
 & Single & \textbf{2}$\times$(\textbf{2}$\times$6) & 76.5 s & 12.79$\times$ & \textbf{2}$\times$(\textbf{64}$\times$6) & 10.0 m & 7.87$\times$ \\
 & Mixed & \textbf{4}$\times$(\textbf{1}$\times$6) & 42.2 s & 23.19$\times$ & \textbf{4}$\times$(\textbf{32}$\times$6) & 4.30 m & 18.19$\times$
\end{tabular}%
}
\begin{tablenotes}
  {\notsotiny
  \item *Total Number of Partitions $=$ \textbf{Batch Nodes} $\times$ (\textbf{Data Nodes} $\times$ Partitions per node).
  Data partitions per node is set to six because each node consists of six GPUs.
  } 
\end{tablenotes}
\vspace{-1em}
\end{table}

\begin{figure}[t!]
    \centering
    \includegraphics[width=\columnwidth]{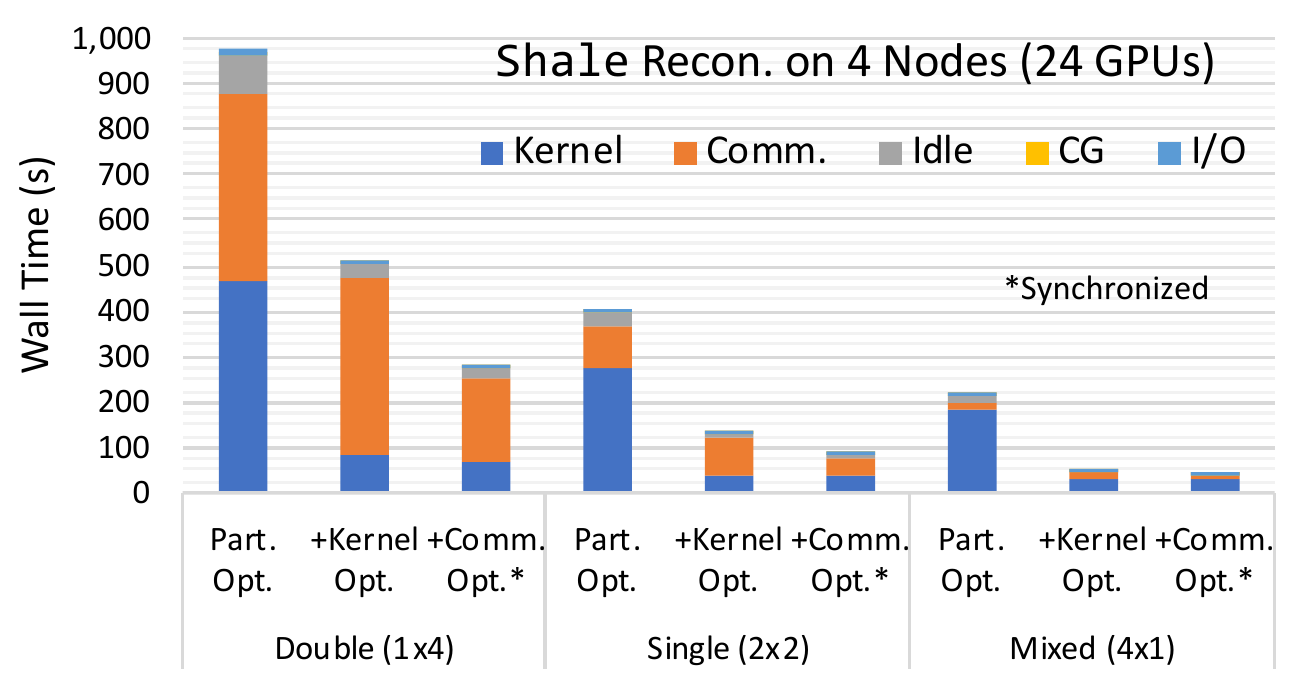}\\
    \small(a)\\
    \vspace{1mm}
    \includegraphics[width=\columnwidth]{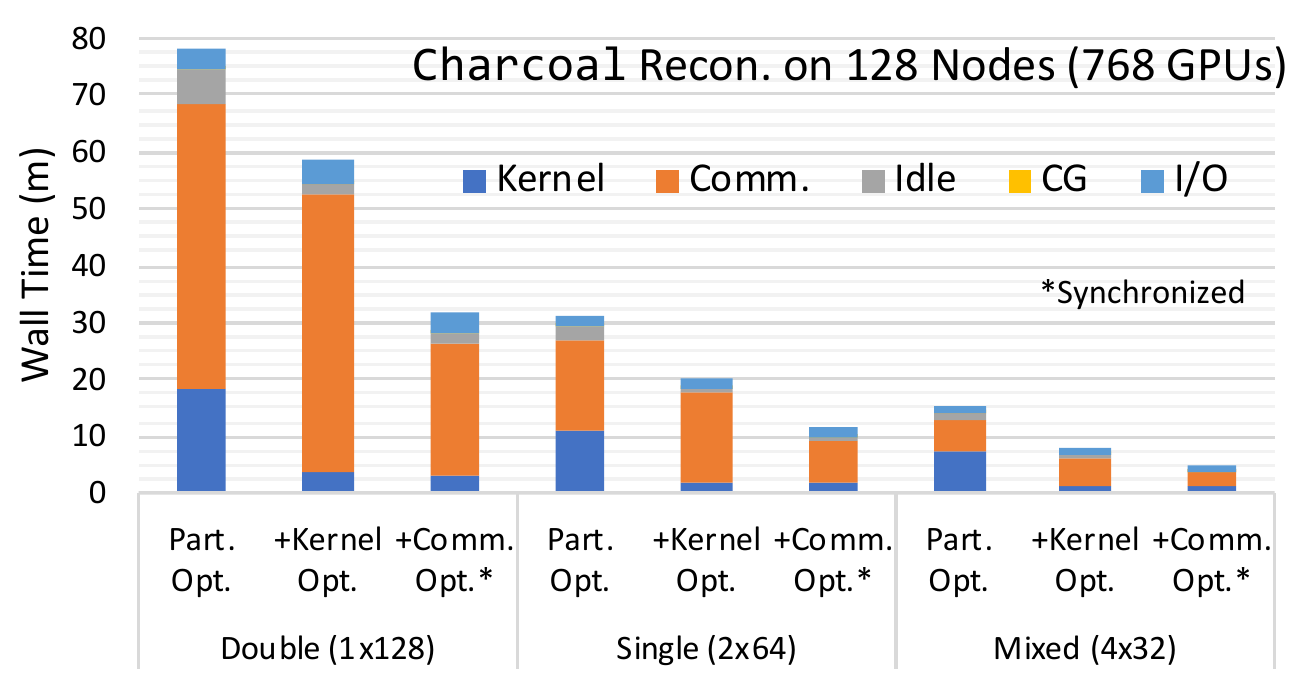}\\
    \small(b)
    \vspace{1mm}
    \caption{Breakdown of end-to-end reconstruction times for (a) \texttt{Shale} and (b) \texttt{Charcoal}. *Overlapping does not applied (i.e., communications are synchronized) to measure each portion's time.}
    \label{fig:overall}
    \vspace{-1em}
\end{figure}

\subsection{Overall Reconstruction Performance}
This subsection evaluates the overall reconstruction speedup with our optimizations. Table~\ref{tab:overall} reports end-to-end reconstruction time for {\tt Shale} on four nodes and {\tt Charcoal} on 128 nodes.
These are the minimum number of nodes required to fit the corresponding memory footprints of double-precision reconstructions. The first three rows report the result without optimized SpMM, hierarchical communications, or communication overlapping. We only optimize the baseline implementation so that it uses the optimal combinations of batch and data parallelism according to the level of precision used for storing the data in memory.   That is, lower precision representations shrink the memory footprint, allowing more batch parallelization and less data partitioning. As explained in Sec.~\ref{sec:optimal}, data partitioning comes with communication overhead and thus it is desirable to employ more batch parallelism and less data partitioning. In this example, as shown in Table~\ref{tab:overall}, we perform partitioning in node granularity, where each node handles six data partitions (one per GPU). The batch nodes partition slices in the 3D reconstruction domain  (data duplication, no communication) and data nodes partition each slice (data partitioning, communication). The next three rows apply kernel optimizations via optimized SpMM implementation (Sec.~\ref{sec:computation}). The last three rows apply hierarchical communications (Sec.~\ref{sec:communication}) and overlapping (Sec.~\ref{sec:overlapping}). As seen in Table~\ref{tab:overall}, overall speedup over double-precision baseline is 23.19$\times$ and 18.19$\times$ for {\tt Shale} and {\tt Charcoal} reconstructions, respectively.

To further study the effect of each optimization, Fig.~\ref{fig:overall} shows the breakdown of the end-to-end reconstruction time. The communications are synchronized (i.e., not overlapped) in order  to accurately measure the time taken by each activity.
However, we do not synchronize GPU kernels and thus idle time reflects the corresponding load imbalance. These results show that optimized SpMM reduces kernel execution time significantly in all cases. The performance 
of the kernel is measured at 75 TFLOPS for \texttt{Shale} on four nodes (24 GPUs) and 2.38 PFLOPS for \texttt{Charcoal} on 128 nodes (768 GPUs). As shown in Fig.~\ref{fig:overall}, execution time is dominated by communication for most of the cases. Hierarchical communications reduce the communication time by more than 50\% in all cases. Sec.~\ref{sec:kernel_performance} and Sec.~\ref{sec:communication_performance} further analyze the optimized SpMM and communication optimizations in detail.

\subsection{Optimized SpMM Performance}\label{sec:kernel_performance}
Fig.~\ref{fig:speedup}(a) shows optimized SpMM speedup with varied minibatch sizes: Performance increases in all cases with larger minibatching due to the aforementioned register reuse provided by our optimized SpMM. However, the kernel performance stagnates when minibatch size reaches around 16, and then it drops with larger minibatch sizes. This is due to a combination of register pressure and synchronization overhead introduced by large minibatch sizes as discussed in Sec.~\ref{sec:computation}. Nevertheless, minibatch sizes of 18, 28, 16, and 20 provide a maximum of 6.47$\times$, 7.77$\times$, 6.30$\times$, and 6.58$\times$ kernel speedup compared to no minibatching with double, single, half, and mixed precision, respectively. Overall, mixed precision achieves the best performance of 15.66$\times$ speedup over the double precision baseline.

\subsubsection{Roofline Analysis}\label{sec:roofline}
To analyze the optimized SpMM performance further, Fig.~\ref{fig:speedup}(b) shows the roofline analysis plot, where the horizontal axis shows the arithmetic intensity, i.e., the number of floating-point operations per byte accessed from memory, and the vertical axis shows the GFLOPS performance per GPU. Each data point in the figure corresponds to one data point in Fig.~\ref{fig:speedup}(a), where increasing minibatch size increases the arithmetic intensity thanks to the 
data reuse from registers (Sec.~\ref{sec:computation}). Lowering precision naturally increases arithmetic intensity due to the less number of bytes per data element. Half and mixed precisions yield the same arithmetic intensity. The memory bandwidth bound (maximum performance possible) with respect to the theoretical 900 GB/s memory bandwidth of V100 GPU is included in Fig.~\ref{fig:speedup}(b). This figure shows that the performance with small arithmetic intensities is well-bounded by the memory bandwidth and starts to diverge when intensity increases. This is mainly due to the synchronization overhead of multi-stage input buffering. 

Double and half precision performance start to degrade for minibatch sizes larger than 18 because of the register spilling as a consequence of the pressure incurred (see Sec.~\ref{sec:register_pressure}). On the other hand, single and mixed precisions do not experience register spilling up to a minibatch size of 50. This is mainly because register usage is well optimized for single precision unrolled FMAs (Line 26 of Listing~\ref{listing:optimized_spmm}). Nevertheless, performance drops significantly at minibatch sizes of 28 and 20 for single and mixed precisions, respectively. This sharp drop in performance may be due to a change in compiler optimization strategy to prevent register spilling by sacrificing performance under a high register pressure. However, \texttt{nvcc} is proprietary, preventing us from further investigation or verification. 16 is set as the minibatch size for all our experiments since the slices count is often a multiplication of 16.

\begin{figure}[t!]
    \centering
    \includegraphics[width=0.9\columnwidth]{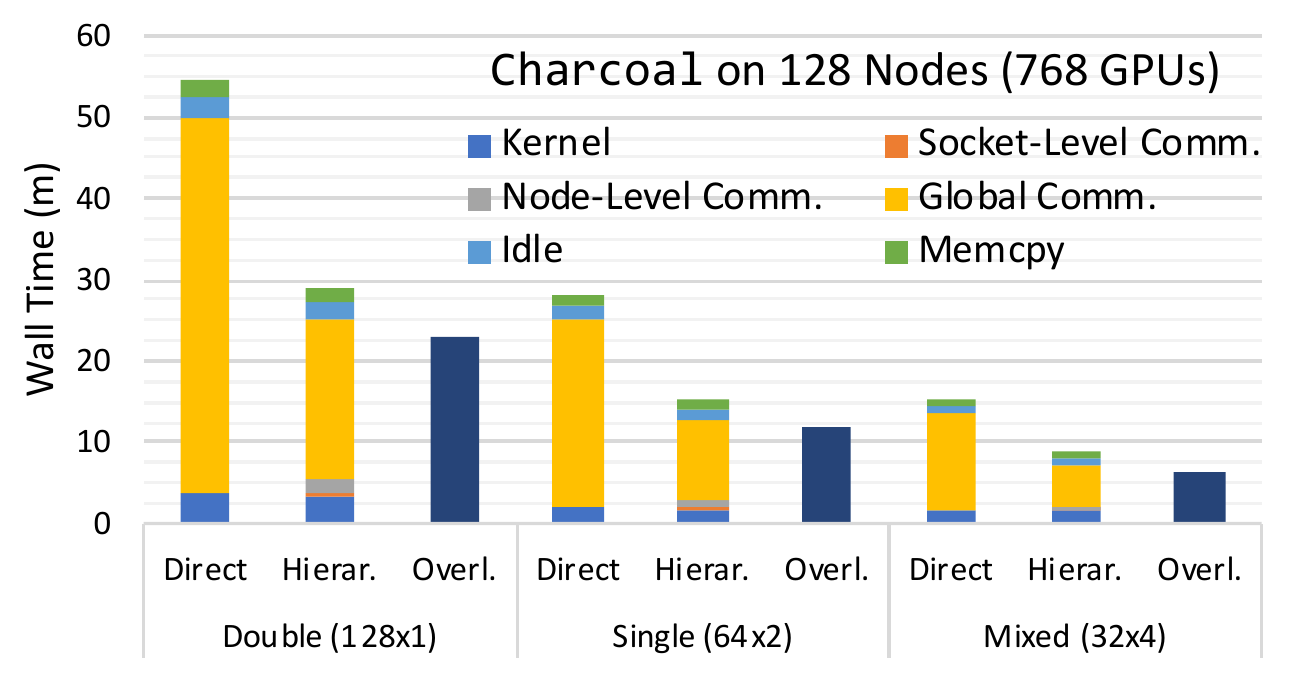}
    \vspace{1mm}
    \caption{Communication time breakdown, \texttt{Charcoal} on 128 nodes.}
    \label{fig:communication}
\end{figure}

\begin{table}[t!]
\vspace{8mm}
\centering
\caption{Communicated Data* and Effective System Bandwidth}
\vspace{1mm}
\label{tab:communications}
\resizebox{1\columnwidth}{!}{%
\begin{tabular}{cc|cc|cc|cc|c}
\textbf{} &  & \multicolumn{2}{c|}{\textbf{Socket-Level Comm.}} & \multicolumn{2}{c|}{\textbf{Node-Level Comm.}} & \multicolumn{2}{c|}{\textbf{Global Comm.}} & \textbf{Memcpy} \\
 & \textbf{Prec.} & \textbf{Data} & \textbf{B/W} & \textbf{Data} & \textbf{B/W} & \textbf{Data} & \textbf{B/W} & \textbf{B/W} \\ \cline{2-9}
\multirow{3}{*}{\textbf{Direct}} & Double & \multicolumn{2}{c|}{} & \multicolumn{2}{c|}{} & 36.6 TB & 1.61 TB/s & 35.2 TB/s \\
 & Single & \multicolumn{2}{c|}{N/A} & \multicolumn{2}{c|}{N/A} & 18.3 TB & 1.61 TB/s & 34.9 TB/s \\
 & Mixed & \multicolumn{2}{c|}{} & \multicolumn{2}{c|}{} & 9.16 TB & 1.59 TB/s & 34.6 TB/s \\ \hline
\multirow{3}{*}{\textbf{Hierar.}} & Double & 36.6 TB & 174 TB/s & 21.4 TB & 21.3 TB/s & 15.2 TB & 1.58 TB/s & 34.9 TB/s \\
 & Single & 18.3 TB & 170 TB/s & 10.7 TB & 22.8 TB/s & 7.58 TB & 1.55 TB/s & 34.5 TB/s \\
 & Mixed & 9.16 TB & 164 TB/s & 5.35 TB & 23.5 TB/s & 3.79 TB & 1.49 TB/s & 33.6 TB/s
\end{tabular}%
}
\begin{tablenotes}
  {\notsotiny
  \item *Per projection (and backprojection). Fig.~\ref{fig:communication} involves 30 projections and 31 backprojections.} 
\end{tablenotes}
\vspace{-3mm}
\end{table}


\subsubsection{Comparison with cuSPARSE}\label{sec:cusparse}

We compare our optimized SpMM performance with \texttt{cusparseSpMM}, which uses cuSPARSE library\footnote{https://developer.nvidia.com/cusparse}. 
In order to find the best performance of both implementations, we aggressively try all minibatch sizes up to 50 and select the best ones. 
Our results show that our optimizations provide 1.53$\times$ to 2.38$\times$ speedup for double and single precision types, respectively. 
Unfortunately, \texttt{cusparseSpMM} is not able to converge to a solution with half precision type. Therefore we cannot make a direct comparison for other types. We are investigating the reason of this.

\subsection{Communication Performance}\label{sec:communication_performance}
To investigate communication optimizations further, Fig.~\ref{fig:communication} shows the breakdown of communication time for \texttt{Charcoal} reconstruction on 128 nodes. Communication time with direct and hierarchical communication strategies (Sec.~\ref{sec:communication}), and total time with communication overlapping (Sec.~\ref{sec:overlapping}) are investigated. Fig.~\ref{fig:communication} shows that hierarchical communication reduces the total communication time by 52\%. Communication overlapping offers an additional speedup of 21\% to 29\% in total execution time. As opposed to the example in Fig.~\ref{fig:overlapping}, \texttt{Charcoal} reconstruction has less overlapping opportunities 
because the global communication dominates the execution, bounding the overlapping efficiency.

Table~\ref{tab:communications} shows the amount of communicated data in TB and corresponding effective system bandwidth in different levels of the communication hierarchy. Communication data is naturally smaller in all cases with lower precision. Effective system bandwidth is calculated by dividing communication data amount by respective communication time shown in Fig.~\ref{fig:communication}. Table~\ref{tab:communications} shows that the effective bandwidth within each socket is about $100\times$ faster than that among nodes. Similarly, the effective bandwidth among sockets is $15\times$ faster than that among nodes. This high bandwidth within nodes alleviates the local communication overhead for hierarchical communications as seen in Fig.~\ref{fig:communication}. Overall, hierarchical communication reduces the communication among nodes by 58\% for all precision types.

\subsection{Scaling Performance}\label{sec:scaling}
To investigate the scaling properties of the proposed application, we perform two strong and one weak scaling experiments with all optimizations applied. Communication overlapping is disabled to correctly measure individual timings. All experiments perform 30 CG iterations with mixed precision.
\subsubsection{Strong Scaling}\label{sec:strong}
\begin{figure}[h!] 
    \centering
    \includegraphics[width=\columnwidth]{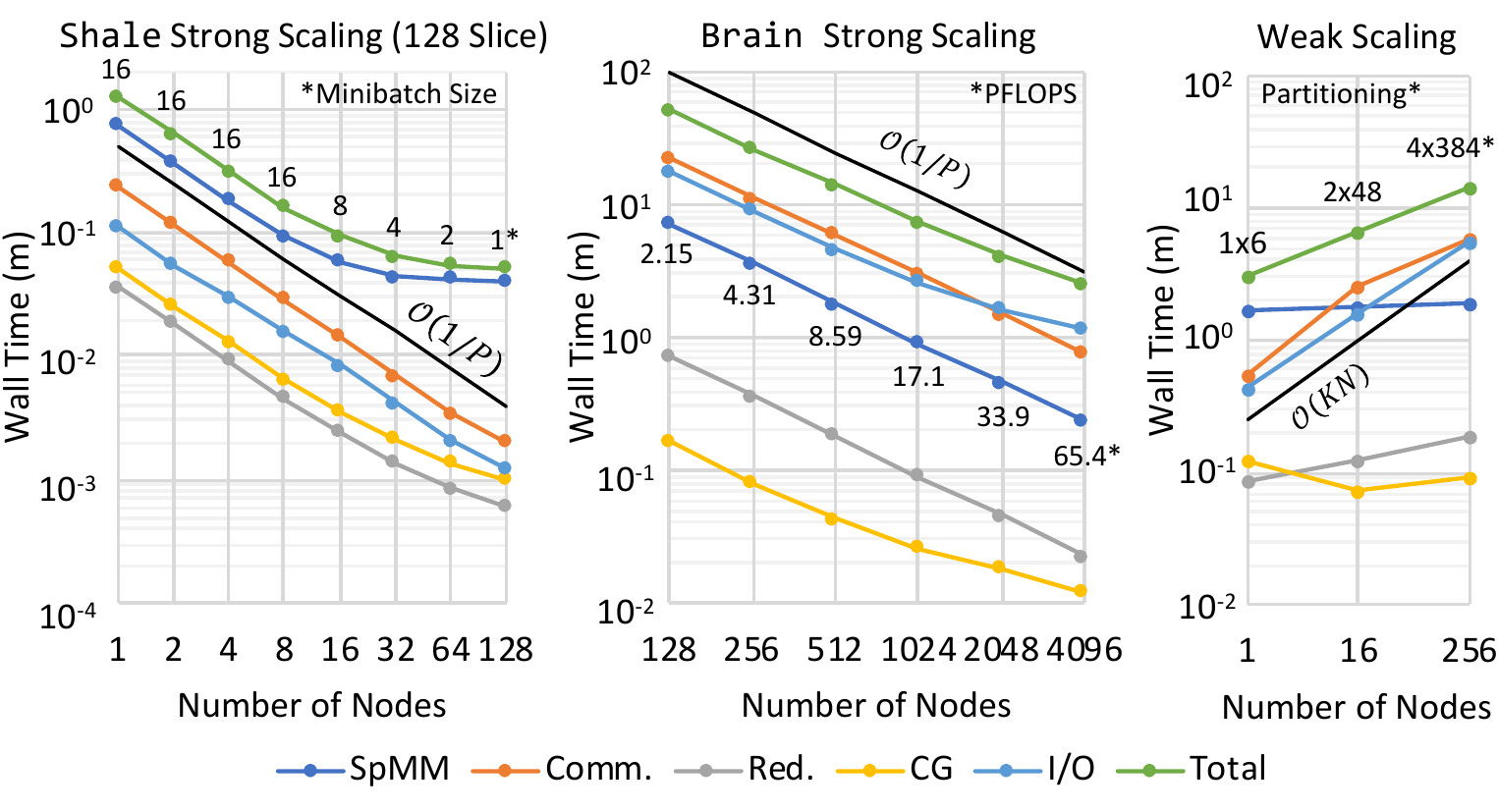}\\
    \vspace{-1mm}
    \hspace{1.1cm}(a)\hspace{2.9cm}(b)\hspace{2.5cm}(c)
    \vspace{1mm}
    \caption{Strong and weak scaling results: Shale and Brain datasets.}
    \vspace{1mm}
    \label{fig:scaling}
\end{figure}

First, we scale 128 slices of \texttt{Shale} up to 128 nodes. Fig.~\ref{fig:scaling}(a) shows the scaling results. This experiment demonstrates the limited scalability when there is a small number of slices. 
This experiment scales up to only 128 nodes where each node reconstructs only one slice. In the base case, there are eight minibatches (each contains 16 slices) so the solution can scale efficiently only up to eight nodes (see~\ref{sec:optimal}). We need to reduce the size of the minibatches after eight nodes to achieve more parallelism by working on finer-granularity. However, SpMM performance drops due to reduced data reuse from registers.
As a result, we scale up to 128 nodes (one slice per node) but the performance is suboptimal due to the bottleneck of reduced SpMM performance as shown in Fig.~\ref{fig:scaling}(a). 

Fig.~\ref{fig:scaling}(b) shows the strong scaling result of \texttt{Brain} reconstruction from 128 nodes (the minimum number of nodes that \texttt{Brain} fits) to \num{4096} nodes (89\% of Summit). As opposed to the previous experiment, there are 9209 slices in \texttt{Brain} (see Table~\ref{tab:dataset}) that allow us to scale without compromising from SpMM performance. As a result, the total reconstruction time is in agreement with $\mathcal{O}(1/P)$ curve, where $P$ is the number of nodes. The I/O performance starts to degrade at 1024 nodes because of the contention that parallel I/O 
puts into the file system. Nevertheless, thanks to our proposed optimizations, {\bf we reconstruct 3D \texttt{Brain} dataset in 2.5 minutes on Summit.} To the best of our knowledge, this is the largest XCT image reconstruction in near-real time. The optimized SpMM kernel performance reaches 65.4 PFLOPS on \num{24576} GPUs (six GPUs per node): 34\% of Summit's theoretical peak performance.

\subsubsection{Weak Scaling}\label{sec:weak}
To investigate the weak scaling properties, we take \texttt{Shale} as the base dataset, and then synthetically double each dimension in the aforementioned measurement data cube ($K$$\times$$ M$$\times$$N$). 
Each time we double all dimensions, and the nominal computation (excluding the parallelization overhead) increases by 16$\times$ (see Table~\ref{tab:complexity}). 
Accordingly, we increase the number of nodes 16$\times$ each time we double measurement dimensions. 
As discussed in Table~\ref{tab:complexity}, the memory footprint increases by 8$\times$ at each step of the scaling. 
We apply the suggested optimal partitioning strategy in Sec.~\ref{sec:optimal} by partitioning data structures among eight nodes and slices between two nodes. 
Fig.~\ref{fig:scaling}(c) shows the weak scaling results. 
Since nominal computation per node remains the same, SpMM time remains constant accordingly. 
However, communication and I/O time increases and becomes the bottleneck for large problems. I/O could be further optimized either by a custom design or a high-performance library, which is beyond the scope of this work.

\subsection{Convergence Performance}\label{sec:convergence}
\begin{figure}[t!] 
    \centering
    \includegraphics[width=\columnwidth]{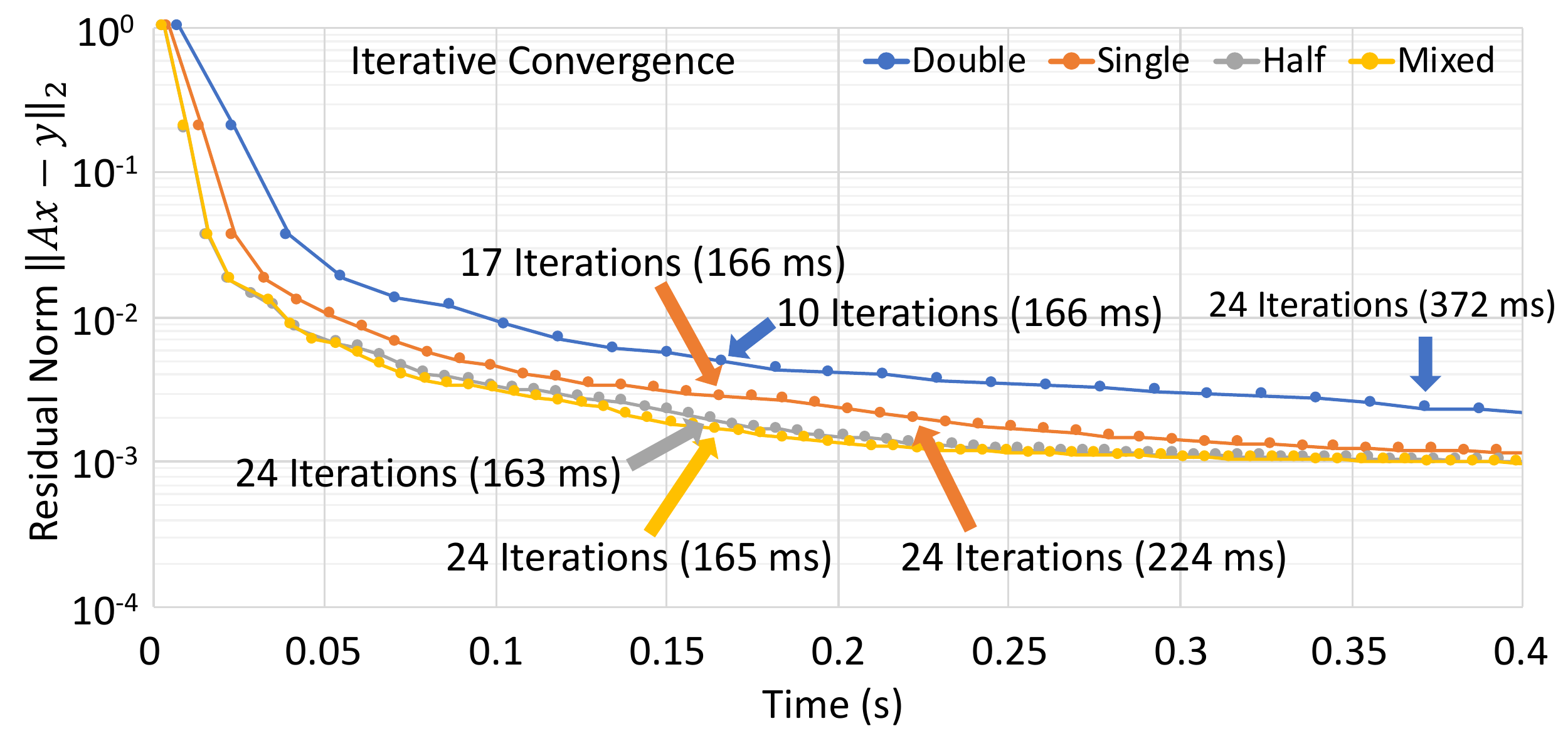}\\
    \vspace{1mm}
    \caption{Convergence speed for \texttt{Chip} with various precisions.}
    \label{fig:convergence}
\end{figure}

To investigate iterative convergence rates, we reconstruct a slice from \texttt{Chip} dataset. Since this dataset is noisy, iterative solution experiences noise \textit{overfitting} after 24 iterations and even though the residual norm shrinks further, measurement noise manifests into the image reconstruction. 
To prevent that, we terminate the iterative solution after 24 iterations. 
No serious convergence problem is observed with reduced precisions. 
This is mainly because the numerical noise floor is \reb{well below} the measurement noise level---this applies to other datasets as well. 
Fig.~\ref{fig:convergence} shows the (relative) residual norm for a single slice with respect to the execution time with double, single, half, and mixed precisions. 
Mixed (and half) precision performs 24 iterations in 165 ms, while single- and double-precision implementations complete the same number of iterations with 224 and 372 ms, respectively.

\section{Related Work}\label{sec:related}
\hspace{-2mm}
Tomographic reconstruction has been researched extensively over the years. 
Iterative reconstruction algorithms usually show better image quality compared to analytical approaches~\cite{bicer2017trace, aditya:timbir, venkatakrishnan2013plug, bruyant2002analytic, fessler2000statistical} and have been used to accommodate limitations on experimental dataset, e.g. noisy measurements, missing angles and others~\cite{aslan2019joint, nikitin2019photon, ching2018rotation}. 

Many parallelization techniques have been developed to ease the computational requirements of iterative algorithms, including distributed computing methods using multi-core~\cite{treibig2012pushing, agulleiro2011fast, jones:hybrid_recon:NC06, johnson:data_parallel_iter:MPC99}. However, they found limited applicability for large-scale dataset due to the computational requirements. 

Many-core architectures, such as GPUs, have been widely used for reconstruction small to medium dataset~\cite{sabne2017model, van2015astra, li2018cumbir} and received significant attention in recent years \cite{hidayetouglu2019memxct, yu2019gpu, wang2019consensus, chen2019ifdk, hidayetoglu2018}.
Medical imaging is one of the areas that extensively utilizes advanced tomographic reconstruction techniques. 
Since limiting dose exposure to patients is crucial, iterative reconstruction approaches are widely used to accommodate noisy measurements~\cite{Lee:Neuro_Imaging_CUDA:CMPB12, Chou:Medical_Imaging_GPUs:MP11, Jang:MultiGPU_Iterative_Recon:BI09}.
Many-core systems can deliver the computational throughput required by the reconstruction tasks, however their limited memory and (host-device) communication cost can introduce significant overhead~\cite{jablin2011automatic}.
In contrast, our solution provides mixed precision computations and multi-level reduction techniques to ease communication cost.


Synchrotron radiation facilities can generate small to extreme-scale data using variety of imaging modalities, including tomography~\cite{ gursoy2015hyperspectral, duke2016time, gursoy2015maximum}. 
High performance software infrastructures have been built to handle tomographic reconstruction workflows that require large-scale compute resources~\cite{bicer2017real, pandolfi2018xi, bicer2016optimization, peterka2016diy2, yildiz2019, liu2019deep}. 
The end-to-end performance of these systems heavily relies on underlying reconstruction engines and can directly benefit from the optimizations proposed in our work.

\section{Conclusion}\label{sec:conclusion}

In this work, we introduce novel optimization techniques for MemXCT approach that exploit 3D volume properties and multi-GPU node architecture during (back)projection operators.
Specifically, our simultaneous data and batch partitioning scheme provides configurable volume distribution among processes and enables reconstruction of extremely large tomography data; 
XCT-Optimized SpMM design considers the sparse matrix structure and efficiently provides GPU utilization by reusing data from shared-memory and registers; hierarchical communications performs extra intra-node communications to minimize the inter-node communication; and finally, effective use of mixed-precision types further reduces memory footprint and communication volume while maintaining the reconstruction quality. The algorithm design and optimizations described in this paper enable an 11k$\times$11k$\times$9k 3D brain image reconstruction under three minutes using 24,576 GPUs on Summit supercomputer, reaching 65 PFLOPS: 34\% of Summit’s peak performance. 


\section*{Acknowledgments}

This material was partially supported by the U.S. Department of Energy, Office of Science, Advanced Scientific Computing Research and Basic Energy Sciences, under Contract DE-AC02-06CH11357. This research used resources of the Oak Ridge Leadership Computing Facility at the Oak Ridge National Laboratory, which is supported by the Office of Science of the U.S. Department of Energy under Contract No. DE-AC05-00OR22725. We thank Narayanan (Bobby) Kasthuri from UChicago/Argonne, and Rafael Vescovi and Ming Du from Argonne for sharing the mouse brain dataset. 
The mouse brain, IC chip, and activated charcoal data were collected by Vincent De Andrade at beamline 32-ID, Advanced Photon Source, Argonne National Laboratory.
This work is supported by IBM-ILLINOIS Center for Cognitive Computing Systems Research (C3SR).
This research is based in part upon work supported by the Center for Applications Driving Architectures (ADA), a JUMP Center co-sponsored by SRC and DARPA. 
This material is supported by  funding from the European Union's Horizon 2020 research and innovation programme under the Marie Skłodowska-Curie grant agreement H2020-MSCA-COFUND-2016-754433.

\bibliographystyle{ieeetr}
\bibliography{bibliography,bicer,recon}



\end{document}